\documentclass[amsmath,amssymb,aps,pre,groupedaddress,superscriptaddress,twocolumn]{revtex4-2}
\usepackage[utf8]{inputenc}
\usepackage[english]{babel}
\usepackage[english]{varioref}
\usepackage{graphicx}
\usepackage{dcolumn}
\usepackage{bm}
\usepackage{amsmath}
\usepackage{amssymb}
\usepackage{caption}
\usepackage[colorlinks=true,linkcolor=blue,urlcolor=blue,citecolor=blue,anchorcolor=blue]{hyperref}

\begin{document}

\title{Preserving system activity while controlling epidemic spreading \\in adaptive temporal networks}

\author{Marco Mancastroppa}
\affiliation{Aix Marseille Univ, Université de Toulon, CNRS, CPT, 13009 Marseille, France}
\author{Alessandro Vezzani}
\affiliation{IMEM-CNR, Parco Area delle Scienze 37/A 43124 Parma, Italy}
\affiliation{Dipartimento di Scienze Matematiche, Fisiche e Informatiche, Università degli Studi di Parma, Parco Area delle Scienze, 7/A 43124 Parma, Italy}
\affiliation{INFN - Istituto Nazionale di Fisica Nucleare, Gruppo Collegato di Parma, Parco Area delle Scienze 7/A, 43124 Parma, Italy}
\author{Vittoria Colizza}
\affiliation{INSERM - Institut national de la santé et de la recherche médicale, Sorbonne Université, Institut Pierre Louis d'Epidémiologie et de Santé Publique (IPLESP), 75012 Paris, France}
\affiliation{Department of Biology, Georgetown University, Washington, District of Columbia, USA}
\author{Raffaella Burioni} 
\affiliation{Dipartimento di Scienze Matematiche, Fisiche e Informatiche, Università degli Studi di Parma, Parco Area delle Scienze, 7/A 43124 Parma, Italy}
\affiliation{INFN - Istituto Nazionale di Fisica Nucleare, Gruppo Collegato di Parma, Parco Area delle Scienze 7/A, 43124 Parma, Italy}

\begin{abstract}
Human behaviour strongly influences the spread of infectious diseases: understanding the interplay between epidemic dynamics and adaptive behaviours is essential to improve response strategies to epidemics, with the goal of containing the epidemic while preserving a sufficient level of operativeness in the population. Through activity-driven temporal networks, we formulate a general framework which models a wide range of adaptive behaviours and mitigation strategies, observed in real populations. We analytically derive the conditions for a widespread diffusion of epidemics in the presence of arbitrary adaptive behaviours, highlighting the crucial role of correlations between agents behaviour in the infected and in the susceptible state. We focus on the effects of sick-leave, comparing the effectiveness of different strategies in reducing the impact of the epidemic and preserving the system operativeness. We show the critical relevance of heterogeneity in individual behavior: in homogeneous networks, all sick-leave strategies are equivalent and poorly effective, while in heterogeneous networks, strategies targeting the most vulnerable nodes are able to effectively mitigate the epidemic, also avoiding a deterioration in system activity and maintaining a low level of absenteeism. Interestingly, with targeted strategies both the minimum of population activity and the maximum of absenteeism anticipate the infection peak, which is effectively flattened and delayed, so that full operativeness is almost restored when the infection peak arrives. We also provide realistic estimates of the model parameters for influenza-like illness, thereby suggesting strategies for managing epidemics and absenteeism in realistic populations.
\end{abstract}

\maketitle

\section{Introduction}
In recent years, control measures to mitigate the effects of epidemics have become one of the most important topics in the study of epidemic processes \cite{fraser2004,anderson2020individual,prem2020effect,DiDomenico2020,Mancastroppa2021CT,hollingsworth2011mitigation}. It has become increasingly clear that desirable strategies are those that focus on the optimisation of mild, if possible population-targeted, measures that allow to control the epidemic while preserving the essential services and systems of the population, i.e. preserving the system activity and operativeness \cite{Massaro2018,aguilar2021absenteism_strategies,hollingsworth2011mitigation,fenichel2013economic,Thommes2016,bonaccorsi2020}. This often corresponds to the implementation of several layers of interventions (e.g. integrating social distancing with contact tracing), which generally depend on the severity and contagiousness of the epidemic \cite{hollingsworth2011mitigation,Massaro2018,aguilar2021absenteism_strategies,DiDomenico2021NatComm,Mancastroppa2021CT,Mancastroppa2022CT}.

From this perspective, it is crucial to understand how the population operativeness evolves in relation to the epidemic dynamics, as these two dynamics are deeply coupled by adaptive behaviours, i.e. changes in individual behaviour due to the presence of the epidemic \cite{funk2010review,feniche2011adaptive,marceau2010,perra2011Char}. This is fundamental for the management of the epidemic and its consequences, and thus for the design of sustainable response strategies that do not disrupt the system activity and operativeness. For instance, mitigation measures can lead to the isolation of individuals \cite{eames2010impact,vankerckhove2013impact,ariza2018healthcareseeking}, producing a deterioration in the population activity because of increasing absenteeism. The maximum absenteeism level, which corresponds to the minimum of service activity, and the delay between its occurrence and the peak of infection are crucial, especially when the infection reaches its maximum incidence \cite{graitcer2012effects,Thommes2016,gianino2017,Ip2015}. For example, the absenteeism peak can occur earlier than the infection peak, as observed empirically in some systems \cite{bollaerts2010timeliness,besculides2005evaluation,Donaldson2021}. This is useful for the proper functioning of hospital facilities, as healthcare workers are already back at work when the peak of infection occurs, providing sufficient levels of assistance \cite{gianino2017,Ip2015}. Moreover, some control strategies seeking to maintain the operativeness of the network may have counterintuitive effects \cite{gross2006adaptive,scarpino2016effect,mancastroppa2020active}. For example, one way to maintain the essential services would be to replace infected individuals with new active (and susceptible) ones, e.g. substitute teachers: while these replacement strategies maintain good activity levels, they may accelerate the spread of the epidemic \cite{scarpino2016effect}.

Hence, it is of great importance to develop models that describe the spread of epidemics, accounting for the coupling between the epidemic, the adaptive behaviours and the population operativeness. In building such models, an important point is that interactions between individuals evolve over time \cite{holme2013temporal,masuda2016guide,rodriguez2022} and their dynamics can change during the epidemic due to adaptation \cite{funk2010review,feniche2011adaptive,marceau2010}: for example, typically a population reduces its activity in response to an epidemic but always remains partly active \cite{eames2010impact, vankerckhove2013impact,ariza2018healthcareseeking} and thus capable of sustaining the spread of the epidemic, even under conditions of complete closure (e.g. as seen during the lockdowns in the COVID19 pandemic \cite{DiDomenico2020}). These properties can be effectively described through adaptive temporal networks \cite{holme2013temporal,masuda2016guide,satorras2015epidemic,berner2023adaptive}. Another important feature is that the temporal networks of social interactions are often heterogeneous both structurally and temporally: for example, individuals have different propensities to engage in social interactions, and structurally have largely varying numbers of social contacts over time \cite{mossong2008social,peters2014relative,apolloni2013age,ubaldi2016asymptotic,ribeiro2013quantifying}. Furthermore they can participate in social interactions through highly heterogeneous temporal patterns, both in their formation dynamics and in their duration \cite{karsai2018bursty,ubaldi2017burstiness}. This heterogeneity plays a key role in the spread of epidemics and in their control \cite{vespignani2002,satorras2015epidemic,mancastroppa2019burstiness}. Therefore, models need to account for the network evolution and heterogeneity to deeply understand the impact of mitigation strategies, which must be modulated and adapted to the network features.

In this paper, we formulate a general framework to model a wide range of different adaptive behaviours and mitigation strategies, observed and implemented in real populations exposed to epidemics. We describe the social interactions through activity-driven temporal networks \cite{perra2012activity,liu2014controlling,mancastroppa2020active}, a class of evolving networks where each node is characterized by its activity, i.e. its propensity to establish links with other individuals \cite{perra2012activity}, and the epidemic through the SIS and SIR epidemic processes. To model the co-evolution with the epidemic, the activity $a$ of each node depends on its health status, $(a_S,a_I)$ in the susceptible $S$ and in the infected state $I$, and these are drawn from a joint distribution $\rho(a_S,a_I)$. This approach allows us to model a large set of realistic adaptive behaviours and containment strategies, ranging from "disease-parties" \cite{swine2009,covid_2,pox} to population-targeted mitigation measures: indeed, the functional form of the joint distribution defines the most general adaptive behaviour. We analytically derive the epidemic threshold as a function of the joint distribution $\rho(a_S,a_I)$, and we show the crucial role of the correlations between the nodes behaviour in the infected and in the susceptible state.

We focus on the adaptive behaviour corresponding to sick-leave \cite{eames2010impact,vankerckhove2013impact,ariza2018healthcareseeking}, which consists in activity reduction due to illness or mitigation measures, e.g. social distancing. We compare the effectiveness of different sick-leave strategies in reducing the impact of the epidemic while preserving the system activity. We study both the change in the epidemic threshold and the dynamics of the active phase, and we show that network heterogeneity plays a key role in shaping the most effective adaptive policy. In homogeneous networks all strategies are equivalent and poorly effective, requiring the implementation of strong interventions which disrupt the population operativeness. In heterogeneous networks targeted strategies (over the most vulnerable nodes) are significantly effective, even for small fractions of sick-leaving nodes, i.e. for mild measures, and also maintain a low level of absenteeism and a high system activity. Interestingly, targeted strategies are effective in flattening and delaying the infection peak, and also in anticipating both the maximum of absenteeism and the minimum of population activity, so that the population is almost fully operative during the infection peak. Instead, uniform strategies require strong interventions, producing a sharp reduction in the system activity simultaneous with the infection peak. 

\section{SIS model on adaptive activity-driven networks}\label{sez:model}
We consider an adaptive activity-driven network \cite{perra2012activity}, introducing behavioural changes in the network evolution due to the spread of an epidemic. There are some attempts in this direction in the literature \cite{moinet2018effect,rizzo2014effect,zino2020analysis,parino2021,HOU2023127850}: for example, in \cite{moinet2018effect} an activity-driven model is proposed that includes awareness of the epidemic to mimic the precautions taken by individuals to reduce their risk of infection. A uniform deterministic rescaling of the activity of infected nodes due to infection is proposed in \cite{rizzo2014effect,zino2020analysis}. However, in realistic populations, nodes do not respond uniformly to infection: their response to the epidemic is uneven, depending on the personal propensity to change behavior during the epidemic \cite{ariza2018healthcareseeking,anderson2020individual}.

Our model takes into account this kind of heterogeneous behavioural response, modelling a change in the activity of infected nodes: the activity of a node when healthy and during the infection are extracted from a joint probability distribution, whose properties and functional form define the modelled adaptive behaviour. This approach is quite general and models a wide class of adaptive behaviours, including the homogeneous deterministic rescaling \cite{rizzo2014effect} as a special case, and includes heterogeneous adaptive behaviours in the population. Furthermore it is able to model realistic behaviours detected in populations actually exposed to epidemics, such as \textit{"sick-leave"} \cite{eames2010impact,vankerckhove2013impact,ariza2018healthcareseeking}. The adaptive behaviour considered in our model targets only infected individuals and models a change in activity due to illness (sickness behaviour) or in general social distancing and mild-to-moderate control interventions implemented to contain or mitigate the spread of infectious diseases \cite{anderson2020individual,fraser2004}. The considered behavioural change impacts the way nodes activate links, but does not affect the way in which they receive links, modelling a minimum level of interactions which cannot be completely cancelled \cite{eames2010impact,vankerckhove2013impact,DiDomenico2020}. This marks the difference with the model proposed in \cite{mancastroppa2020active}, where an attractiveness parameter was introduced to model stronger control measures which also impact the contacts passively received by the nodes. Thus, in the proposed model the population preserves an overall activity, mainly driven by healthy nodes as expected for mild-to-moderate control interventions (see Section \ref{sez:rel}); while in strong interventions even the healthy nodes are affected by containment, producing a strong deterioration in the population operativeness \cite{mancastroppa2020active,hollingsworth2011mitigation,fenichel2013economic}. \\

Here we consider the SIS epidemic model: the network is composed of $N$ nodes, and each node $i$ is characterized by two parameters $(a_S^i,a_I^i)$, respectively the activity of node $i$ in the susceptible state $S$ and in the infected state $I$, drawn from the joint distribution $\rho(a_S,a_I)$. Initially all the nodes are disconnected, then the network evolves over time with a continuous dynamics of link formation: each node is characterized by a Poissonian activation dynamics, with rate $a_S$ if the node is susceptible and with rate $a_I$ if the node is infected. When a node activates, it creates one link with $m$ randomly chosen nodes (hereafter we will consider $m=1$ with no loss of generality). If the nodes involved in a contact are one infected $I$ and one susceptible $S$, the susceptible node gets infected with probability $\lambda$ (infectious contact), $S+I \xrightarrow[]{\lambda} 2I$, otherwise
nothing happens during the contact. Then the contact is removed. Infected nodes recover with rate $\mu$, through a Poissonian spontaneous recovery process: $I \xrightarrow[]{\mu} S$. 

Note that a susceptible node activates contagious links with infected nodes with an overall rate $\lambda a_S$, and an infected node creates contagious contacts with susceptible nodes with an overall rate $\lambda a_I$. The formulation of the model is explicitly designed to separate the description of the epidemic dynamics from that of the network dynamics: nodes activate links over time with a continuous dynamic of rate $(a_S,a_I)$, respectively when susceptible and when infected; some of these links transmit the infection (with probability $\lambda$), while others do not. This choice is crucial for monitoring both the evolution of the epidemic and the evolution of the population operativeness (see Section \ref{sez:rel}), by considering the whole set of interactions generated over time, even if they are not epidemiologically relevant. In fact, the population operativeness is measured by the total system activity, which consists of the full set of the interactions generated, whether they transmit the infection or not.

The SIS model presents a phase transition between an absorbing state and an active phase: the control parameter is $r=\lambda \overline{a_S}/\mu$ (effective infection rate) and its critical value is the epidemic threshold $r_C$. If $r<r_C$ the infection reaches only a negligible fraction of nodes and dies out (absorbing phase), while if $r>r_C$ the epidemic produces a large-scale diffusion in the population (active phase). This threshold behavior can also be equivalently formulated in terms of the basic reproduction number $R_0$, that is the average number of secondary infections produced by an individual in a fully susceptible population \cite{satorras2015epidemic}: if $R_0<1$, the epidemic rapidly dies out, while if $R_0>1$, the epidemic propagates and reaches the active diffusion. The critical threshold in this terms is $R_0^C=1$ \cite{satorras2015epidemic}.\\

This simple mean-field model, which neglects memory effects \cite{tizzani2018memory,karsai2014time,ubaldi2016asymptotic,Sun2015cont}, temporal heterogeneities \cite{mancastroppa2019burstiness,ubaldi2017burstiness} and correlations in the activation dynamics, can be analytically solved. In particular an activity-based mean-field approach (ABMF) is exact for the present system given that connections are continuously reshuffled over time, destroying local correlations. Considering $P_{a_S,a_I}(t)$, the probability that a node of activities $(a_S,a_I)$ is infected at time $t$, the following equation for the epidemic dynamics holds:
\begin{widetext}
\begin{equation}
\partial_t P_{a_S,a_I}(t)=-\mu P_{a_S,a_I}(t) + \lambda (1-P_{a_S,a_I}(t))\left[ a_S \int da_I' \int da_S' \rho(a_S',a_I') P_{a_S',a_I'}(t) + \int da_I' \int da_S' \rho(a_S',a_I') a_I' P_{a_S',a_I'}(t) \right],
\label{eq:dynamics}
\end{equation}
\end{widetext}
where the first term in the right-hand side accounts for the recovery process and the second term for contagion processes. Through the ABMF approach we obtain the probability for a node $(a_S,a_I)$ to be infected in the steady state (asymptotic behaviour) $P_{a_S,a_I}^0=\lim\limits_{t \to \infty} P_{a_S,a_I}(t)$ and the epidemic prevalence (order parameter of the phase transition) $\overline{P}= \int da_{S} \int d a_I \rho(a_S,a_I) P_{a_S,a_I}^0$ (see Appendix A):
\begin{equation}
P_{a_S,a_I}^0=\frac{a_S \overline{P}+\overline{a_IP}}{\mu/\lambda + a_S\overline{P}+ \overline{a_I P}},
\label{eq:prevb}
\end{equation}
\begin{equation}
\overline{P}=\frac{\overline{a_I P}}{\mu/\lambda - \overline{a_S} + \overline{a_S P}+ \overline{a_I P}},
\label{eq:prev}
\end{equation}
where $\overline{g(a_S,a_I) P} = \int da_S \int d a_I \rho(a_S,a_I) P_{a_S,a_I}^0 g(a_S,a_I)$ and $\overline{g(a_S,a_I)} = \int da_S \int da_I \rho(a_S,a_I) g(a_S,a_I)$. Considering the asymptotic steady state and applying a linear stability analysis around the absorbing state we obtain the epidemic threshold (see Appendix A for the detailed derivation):
\begin{equation}
r_C= \frac{2 \overline{a_S}}{\overline{a_S} + \overline{a_I} + \sqrt{(\overline{a_S} - \overline{a_I})^2 + 4 \, \overline{a_I a_S}}}.
\label{eq:thr}
\end{equation}
The epidemic threshold $r_C$ strongly depends on the correlations between $a_I$ and $a_S$, captured in $\overline{a_S a_I}$ and in the joint distribution $\rho(a_S,a_I)$, i.e. it deeply depends on the correlations between the nodes behaviour in the infected and in the susceptible state.
If $a_S$ and $a_I$ are uncorrelated, $\overline{a_I a_S}=\overline{a_I}\cdot \overline{a_S}$, the epidemic threshold is:
\begin{equation}
r_C^{uncorr}=\frac{\overline{a_S}}{\overline{a_S} + \overline{a_I}}.
\end{equation}
Eq.\eqref{eq:thr} indicates that, if the activities $a_I$ and $a_S$ are positively correlated the epidemic threshold is lower than in the uncorrelated case (stronger epidemic); on the contrary for negative correlations the threshold is higher (weaker epidemic). Indeed, fixing $\overline{a_S}$ and $\overline{a_I}$, we obtain $r_C/r_C^{uncorr}>1 \Leftrightarrow \overline{a_I a_S} - \overline{a_I} \,\, \overline{a_S}<0$.
Positive correlations favour the epidemic, since the most active susceptible nodes behave like \textit{hubs} also when infected. We notice that negative correlations are limited by the fact that $a_I \geq 0$ and $a_S \geq 0$, therefore $\overline{a_I a_S} \geq 0$, hence maximally negative correlations are achieved if $\overline{a_I a_S} = 0$ and $\overline{a_I} = \overline{a_S}$. 

However, reasonably infected nodes change their behaviour preserving their inclination to activate, thus realistic strategies feature positive correlations. For example, the non-adaptive (NAD) network corresponds to a perfect correlation between $a_I$ and $a_S$, that is $\rho(a_S,a_I)= \rho_S(a_S) \delta(a_I-a_S)$, where $\delta( \cdot )$ indicates the Dirac delta function. In this case the epidemic threshold is the same obtained in \cite{perra2012activity} (see Appendix A):
\begin{equation}
r_C^{NAD}=\frac{\overline{a_S}}{\overline{a_S} + \sqrt{\overline{a_S^2}}}.
\label{eq:NAD}
\end{equation} 
The NAD case represents the baseline reference scenario to evaluate the performance of different control strategies. Indeed, mitigation measures increase the epidemic threshold $r_C$ with respect to the NAD case, by reducing the social activity of infected nodes (see Eq. \eqref{eq:thr}), thus $r_C^{NAD}/r_C \in [0,1]$. Therefore, when assessing the effect of different adaptive behaviours, we consider the ratio $r_C^{NAD}/r_C$: lower values ($\sim 0$) indicate a stronger impact of the measures on the epidemic, while larger values ($\sim 1$) indicate ineffective mitigation strategies.

Hereafter, we will introduce some realistic response strategies to the epidemic, modifying the perfectly positive correlated NAD case. In particular, we fix the activity distribution of susceptible nodes $\rho_S(a_S)$ and we describe the adaptive behaviour by the conditional probability $\rho_{I|S}(a_I|a_S)$: $\rho(a_S,a_I)=\rho_S(a_S) \rho_{I|S}(a_I|a_S)$. In many real systems it has been observed a broad distribution of $a_S$ \cite{perra2012activity}: heterogeneities in activity distribution account for different jobs/social roles (e.g. teachers, doctors, nurses), which require different number of contacts over time, or for different age groups \cite{apolloni2013age,mossong2008social}. Henceforth we will consider a power-law distribution of $a_S$: $\rho_S(a_S) \sim a_S^{-(\nu+1)}$, with lower cut-off $a_m$ and upper cut-off $a_M=\eta a_m$ \cite{perra2012activity,ubaldi2016asymptotic}, with $\eta \in (1,\infty)$, modelling the presence of strong heterogeneities as observed in many human systems, in which $\nu \sim 0.3-1.5$, featuring large activity fluctuations \cite{ubaldi2016asymptotic,perra2012activity,ribeiro2013quantifying, karsai2014time}. \\

\begin{figure*}
\centering
\includegraphics[width=0.329\textwidth]{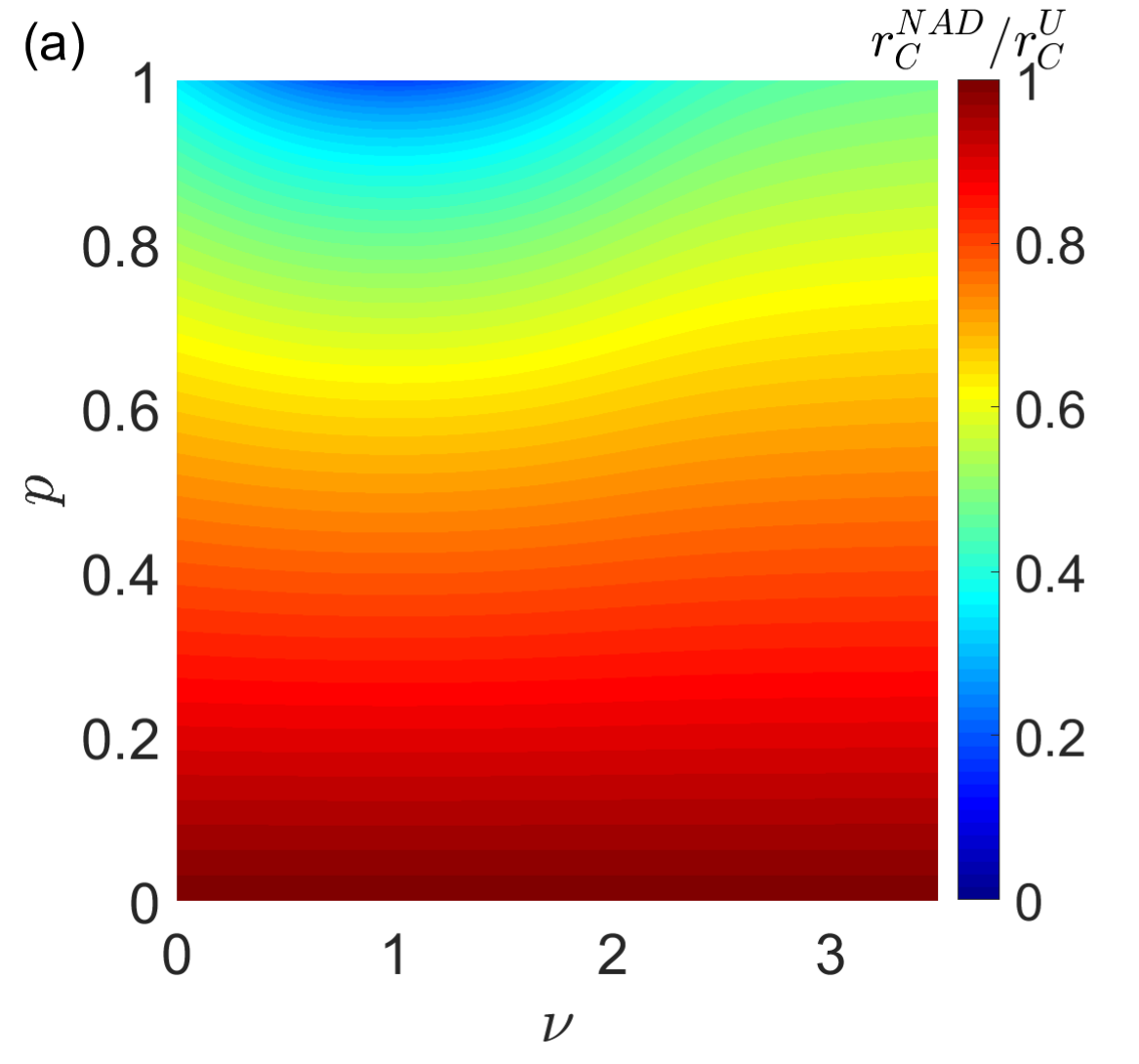}
\includegraphics[width=0.329\textwidth]{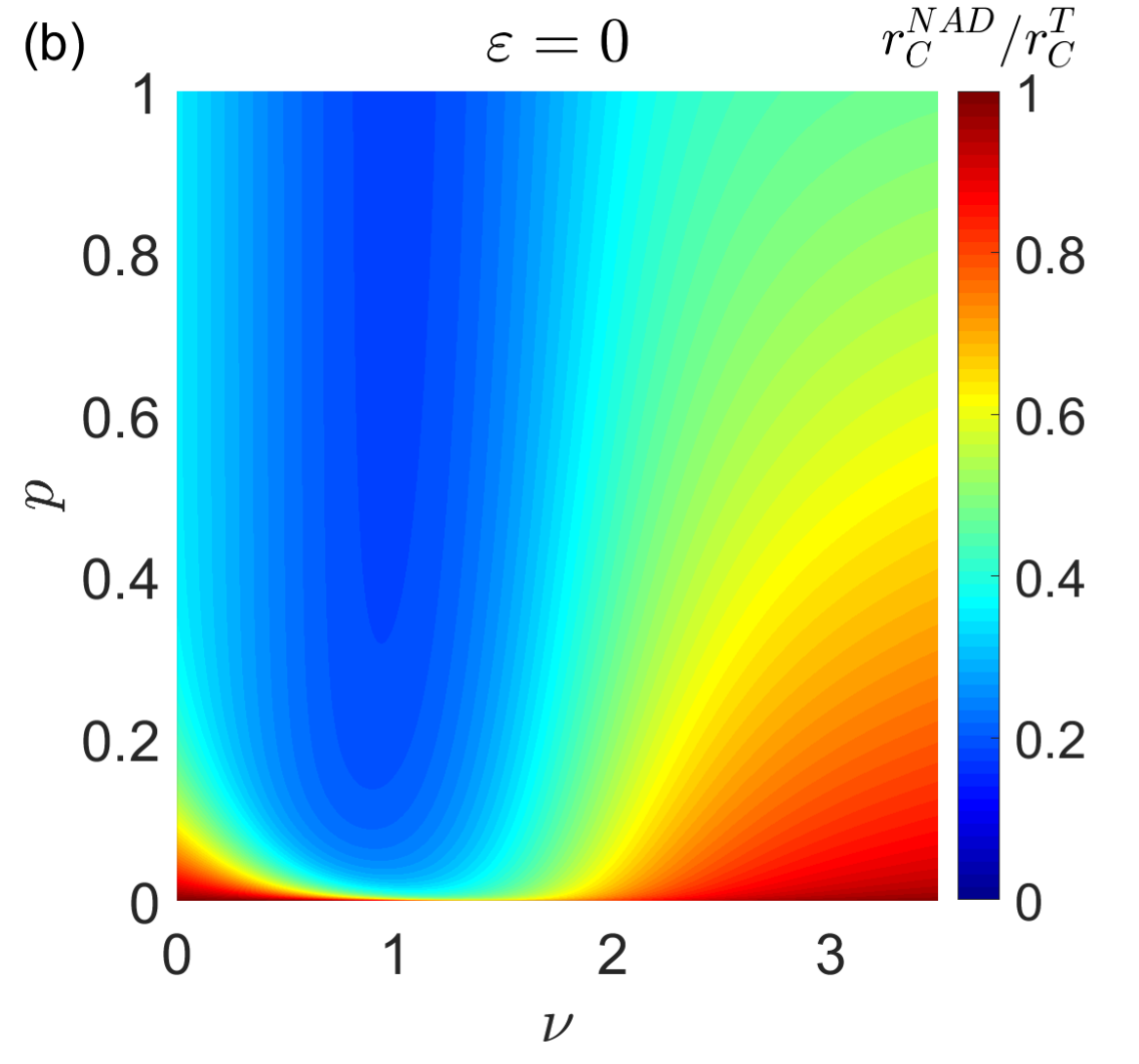}
\includegraphics[width=0.329\textwidth]{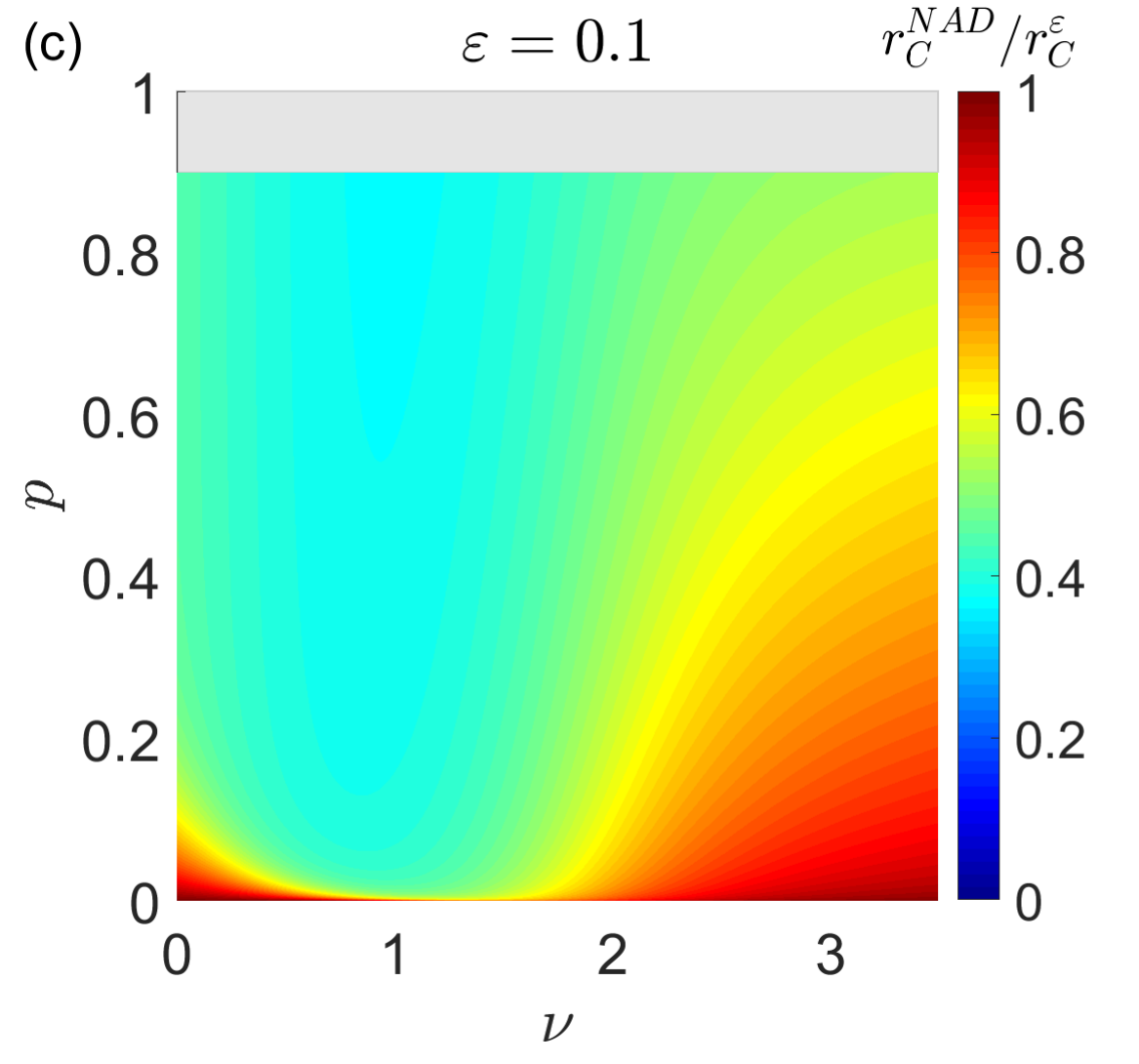}
\caption{\textbf{Epidemic threshold for sick-leave strategies.} In each panel we show the ratio between the epidemic threshold in the NAD (non-adaptive) scenario, $r_C^{NAD}$, and when a sick-leave strategy is implemented, $r_C$. We consider: in panel (a) the uniform strategy, $r_C=r_C^U$; in panel (b) the targeted strategy, $r_C=r_C^T$; in panel (c) the $\varepsilon$-targeted strategy, $r_C=r_C^{\varepsilon}$. The ratio $r_C^{NAD}/r_C$ is plotted as a function of $\nu$, exponent of the $\rho_S(a_S)$ distribution, and $p$, fraction of sick-leaving nodes, through a heat-map. In all cases we fix the distribution of susceptible activity $\rho_S(a_S) \sim a_S^{-(\nu+1)}$ with lower cut-off $a_m=10^{-3}$ and upper cut-off $a_M=1$, moreover in panel (c) we fix $\varepsilon=0.1$, that is the fraction of high-activity nodes not performing sick-leave. Note that the gray zone in panel (c) corresponds to the region of prohibited $p>1-\varepsilon$ values. These results have been obtained analytically from Eqs. \eqref{eq:NAD}, \eqref{eq:rcU}-\eqref{eq:rcE}.
}
\label{fig:comparison_2d}
\end{figure*}

\section{The sick-leave practice}\label{sez:sick}
In many real populations it has been observed an adaptive behaviour that can be described by our model: the \textit{sick-leave} practice \cite{ariza2018healthcareseeking, vankerckhove2013impact,eames2010impact}. A fraction of the population, when infected, cancels its activity $a_I=0$ (sickness induces sick-leave), while the remaining part of the population keeps the same activity in infected and susceptible states $a_I=a_S$ (asymptomatics or paucisymptomatics agents, individuals who must necessarily go to work because of economic reasons or for the preservation of minimum levels of activity in workplaces, such as hospitals, schools and offices). 

In the sick-leave practice we set $\rho_{I|S}(a_I|a_S)= f(a_S) \delta(a_I-a_S) + (1-f(a_S)) \delta(a_I)$, where $\delta( \cdot )$ indicates the Dirac delta function: a node with $a_S$ has a probability $f(a_S)$ to keep its activity when infected, and a probability $1-f(a_S)$ to be inactive when infected ($a_I=0$). The functional form of $f(a_S)$ sets the activity correlations and the sick-leave strategy implemented. The epidemic threshold is (see Appendix B for the detailed derivation):
\begin{widetext}
\begin{equation}
r_C=\frac{2 \overline{a_S}}{\overline{a_S} + \overline{a_S f(a_S)} + \sqrt{\left(\overline{a_S} - \overline{a_S f(a_S)}\right)^2+4 \, \overline{a_S^2 f(a_S)}}}.
\label{eq:rc}
\end{equation}
\end{widetext}
The two limit cases of this approach are $f(a_S)=1 \, \forall a_S$ and $f(a_S)=0 \, \forall a_S$: the former is the NAD network (Eq. \eqref{eq:NAD}), while the latter corresponds to the case in which all nodes perform sick-leave. 

In our model, sick-leaving nodes no longer actively create links, but they passively suffer the activity of the rest of the population, since susceptible and infected nodes are equally contacted by active nodes (even if $a_I=0$). Therefore, every individual engages two types of connections: an active component, which the individual can control and reduce during infection; a passive component which cannot be controlled, due to the rest of the population (e.g. neighbours, doctors, relatives), as observed in real populations exposed to ILI (influenza-like illness) epidemics \cite{eames2010impact, vankerckhove2013impact,DiDomenico2020}. The sick-leave practice is a mild-to-moderate control measure, indeed it does not affect healthy nodes. This marks the difference between \textit{sick-leave} and \textit{quarantine} \cite{mancastroppa2020active}: quarantining nodes would not actively create nor passively receive links. The adoption of quarantine affects also healthy nodes who cannot contact quarantined individuals, deteriorating significantly the population overall activity (strong containment measures). The quarantine can be implemented in our model by introducing an attractiveness parameter \cite{pozzana2017attractiveness,mancastroppa2020active}; if all nodes were in quarantine the epidemic would not spread and the threshold would be infinite \cite{mancastroppa2020active}, on the contrary, if all nodes perform sick-leave the epidemic threshold is still finite $r_C^{best}=1$.

Several works focus on the sick-leave practice in real populations exposed to ILI epidemics \cite{ariza2018healthcareseeking,eames2010impact, vankerckhove2013impact}, providing realistic estimation of our model parameters: for example, in some populations a 75\% reduction in the number of contacts was observed on average in infected nodes \cite{eames2010impact, vankerckhove2013impact}. In \cite{ariza2018healthcareseeking} the behaviour of the French general population during ILI epidemics was investigated, by considering the data collected from the web-based GrippeNet.fr cohort study \cite{grippenet} (which is part of the European consortium InfluenzaNet \cite{influenzanet,influenzanet2}). In particular, considering three influenza seasons between 2012 and 2015, the fraction of the population performing sick-leave $p=\int (1-f(a_S)) \rho_S(a_S) \, da_S$ was estimated \cite{ariza2018healthcareseeking}: about 20\% of the population performs sick-leave when infected, while 50\% does not change its behaviour during the infection and the remaining 30\% instead undertake an intermediate behaviour, reducing the activity but without taking sick-leave. 

We remark that the \textit{sick-leave} practice can be realized in a variety of implementations by appropriately choosing the functional form of $f(a_S)$: in the next Sections we deal with several sick-leave strategies, comparing them and identifying the best ones.

\subsection{Uniform strategy}
The simplest sick-leave strategy is the one acting equally on all the activity classes: every node, independently of its activity $a_S$, has the same probability to take sick-leave. This strategy is expected to be realized in the absence of targeted policies or interventions.  

Hereafter we will refer to it as \textit{uniform strategy}: formally, this case is $ f (a_S) = (1-p) $ $\forall a_S $. A fraction $p$ of the population, when infected, cancels its activity $a_I=0$, while $1-p$ when infected keeps its activity $a_I=a_S$. The threshold is (see Appendix B for the detailed derivation):
\begin{equation}
r_C^{U}= \frac{2 \overline{a_S}}{\overline{a_S} (2-p) + \sqrt{\overline{a_S}^2 p^2 + 4 (1-p) \overline{a_S^2}}}.
\label{eq:rcU}
\end{equation}

The uniform strategy is effective in increasing the epidemic threshold, compared to the NAD case, only when almost all nodes take sick-leave (more than 75\%, $p>0.75$): indeed, the ratio $r_C^{NAD}/r_C^{U}$ is significantly small only when $p \sim 1$ (see Fig. \ref{fig:comparison_2d}(a)).

\begin{figure*}
\centering
\includegraphics[width=0.329\textwidth]{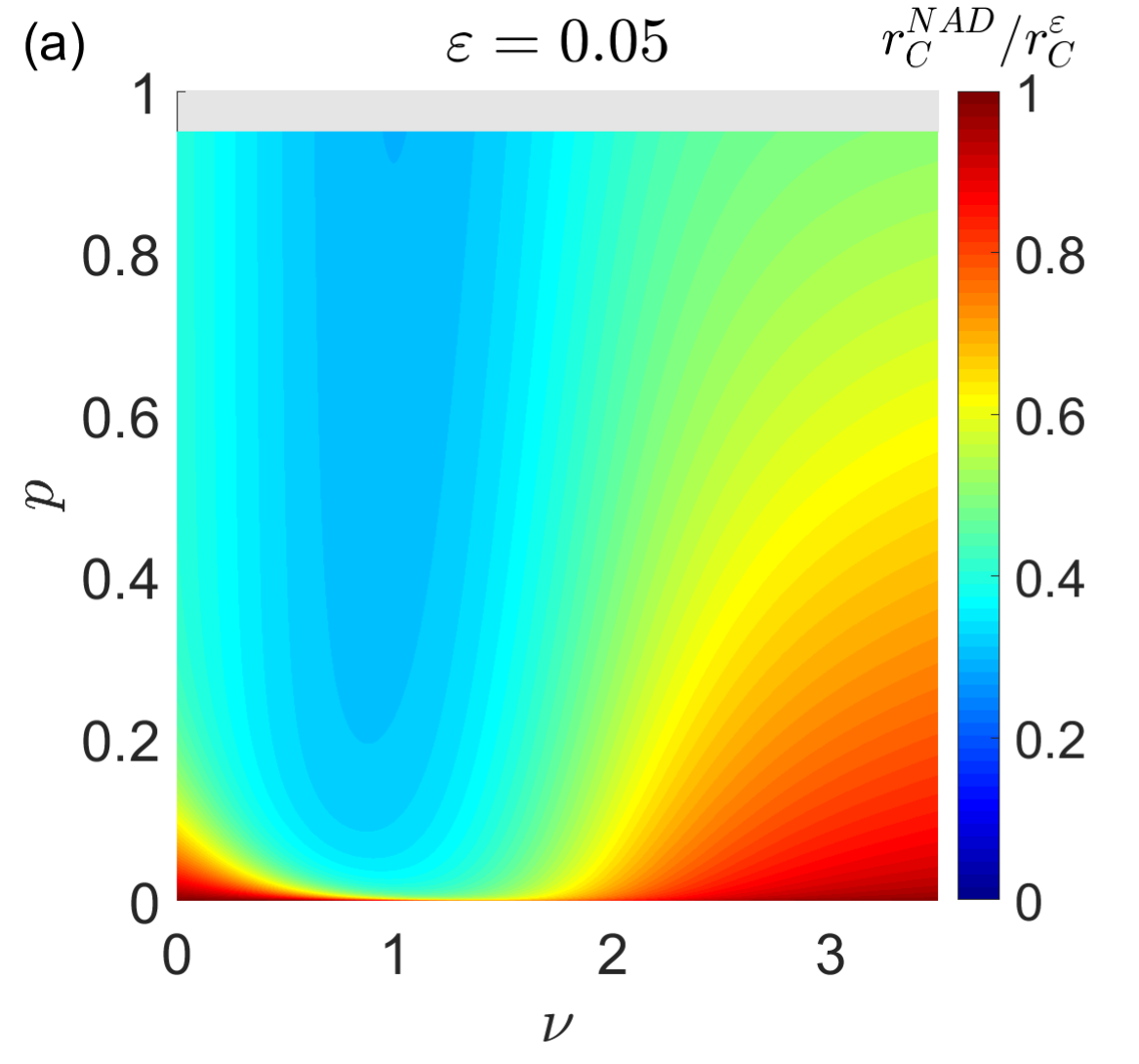}
\includegraphics[width=0.329\textwidth]{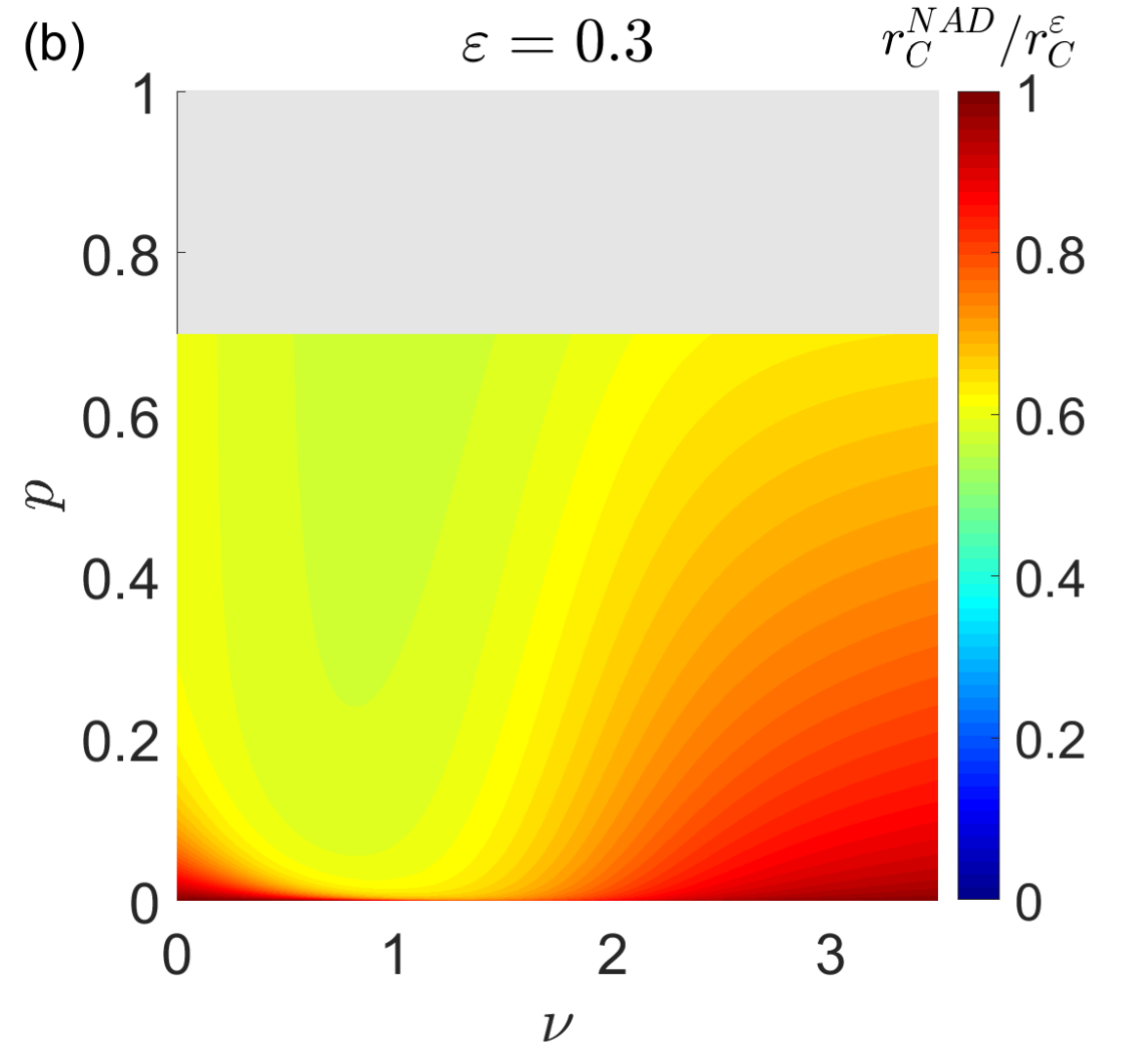}
\includegraphics[width=0.329\textwidth]{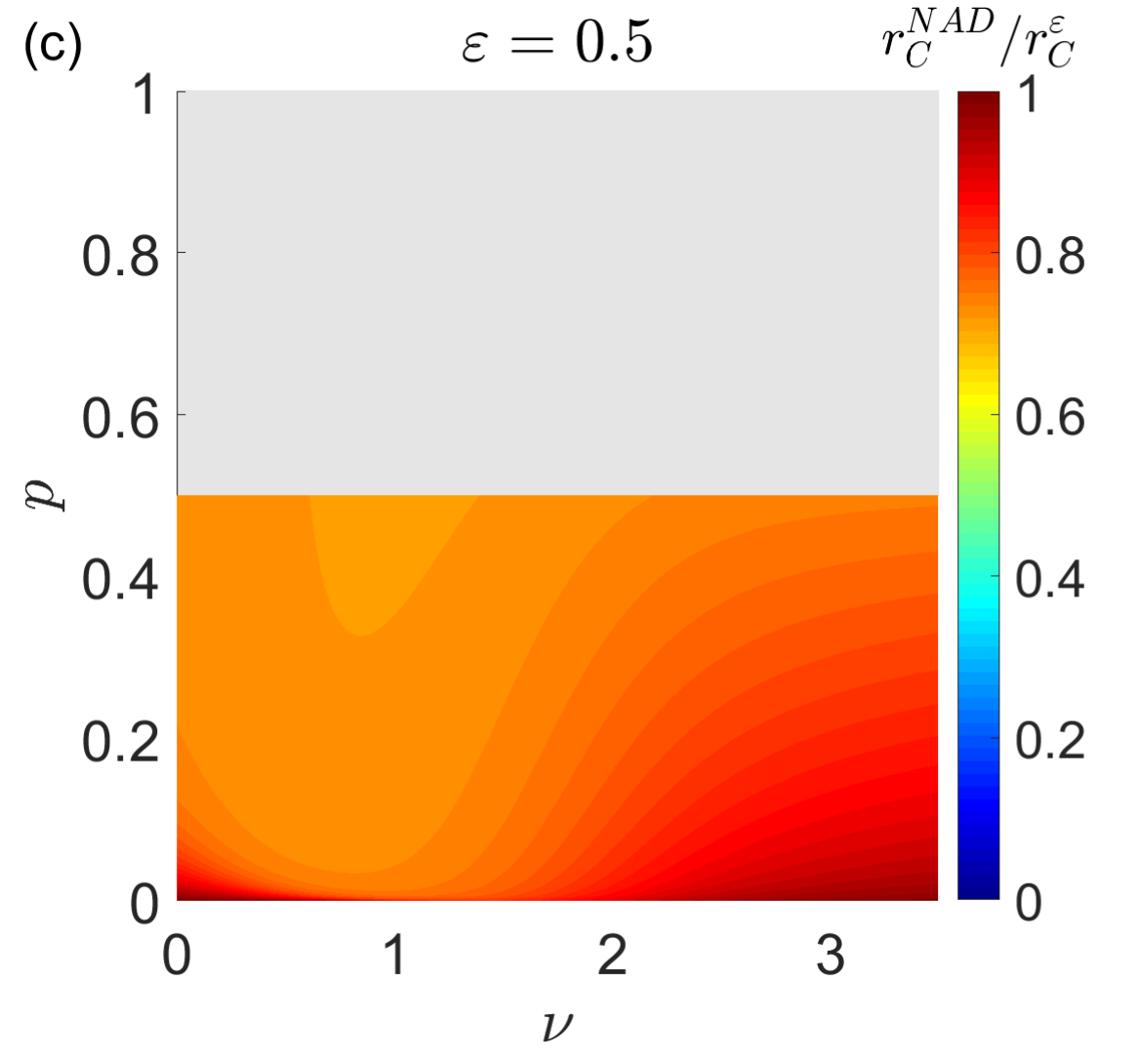}
\caption{\textbf{Epidemic threshold for $\varepsilon$-targeted strategy.} In all panels we consider the ratio between the epidemic threshold in the NAD (non-adaptive) scenario, $r_C^{NAD}$, and when the $\varepsilon$-targeted strategy is implemented, $r_C^{\varepsilon}$. The ratio $r_C^{NAD}/r_C^{\varepsilon}$ is plotted as a function of $\nu$, exponent of the $\rho_S(a_S)$ distribution, and $p$, fraction of sick-leaving nodes, through a heat-map. In all cases we fix the distribution of susceptible activity $\rho_S(a_S) \sim a_S^{-(\nu+1)}$ with lower cut-off $a_m=10^{-3}$ and upper cut-off $a_M=1$. In each panel we consider a different fraction of high-activity nodes not performing sick-leave, $\varepsilon$, reported above each panel. Note that the gray zone in each panel corresponds to the region of prohibited $p>1-\varepsilon$ values. These results have been obtained analytically from Eqs. \eqref{eq:NAD}, \eqref{eq:rcE}.
}
\label{fig:Rapp_epsilon}
\end{figure*} 

\subsection{Targeted strategy}
We consider here a \textit{targeted strategy} over most at risk nodes (high activity $a_S$). In this strategy authorities operate in a selective way, targeting only individuals with high activity when susceptible. The interventions (awareness campaigns and policies) aim that all the nodes with high activity take sick-leave. \\

We set $f(a_S)=\theta(a^*-a_S)$ (where $\theta$ is the Heaviside function and $a^* \in [a_m,a_M]$), in order to consider a completely selective strategy: all nodes with $a_S \geq a^*$ take sick-leave setting $a_I=0$, while nodes with $a_S<a^*$ do not perform sick-leave and keep $a_I=a_S$. In order to make the uniform and targeted strategies comparable, we fix $a^*$ so that in both strategies the fraction of sick-leaving nodes is the same $p$ (see Appendix B). The epidemic threshold is (fixing $y=a^*/a_m$ and $\eta=a_M/a_m$ - see Appendix B for the analytical derivation):
\begin{widetext}
\begin{equation}r_C^T= \frac{2 \overline{a_S}}{\overline{a_S} \left( 1+\frac{1-y^{1-\nu}}{1-\eta^{1-\nu}}\right)+\sqrt{\overline{a_S}^2 \left( \frac{y^{1-\nu}-\eta^{1-\nu}}{1-\eta^{1-\nu}} \right)^2+4 \, \overline{a_S^2} \frac{1-y^{2-\nu}}{1-\eta^{2-\nu}}}}.
\label{eq:rcT}
\end{equation}
\end{widetext}

In networks with homogeneous activities ($\nu \gtrsim 2.5$), a high fraction of sick-leaving nodes ($p>0.75$) is necessary to obtain a significant increase in the epidemic threshold; on the contrary in heterogeneous systems ($\nu \sim 0.3-1.5$) the targeted strategy is effective for almost any value of $p>0.05$ (see Fig. \ref{fig:comparison_2d}(b)). In a heterogeneous network, the same gain in the epidemic threshold can be obtained by imposing a very small (10\%) or a higher number ($\sim$ 100\%) of individuals taking sick-leave: since all the key nodes in the spread of epidemics are already inactive with small $p$. This suggests that the authorities can intervene with low intensities using the targeted approach and achieve a significant reduction in the epidemic threshold.

The targeted strategy is much more effective than the uniform one for heterogeneous activities while for homogeneous activities the two strategies are equivalent and both ineffective (compare panels (a) and (b) of Fig. \ref{fig:comparison_2d}): in the uniform case the region with very low ratio $r_C^{NAD}/r_C^{U}$ is small and located at $\nu \sim 1$ and $p>0.95$; in the targeted case it grows impressively and becomes much deeper, extending in $\nu \sim 0.1 - 1.5$ and $p \sim 0.05 - 1$, making this strategy much more effective for heterogeneous activities. \\ 

\subsection{$\varepsilon$-targeted strategy}
In realistic population it is difficult to impose to each node with $a_S \geq a^*$ to perform sick-leave, since some nodes might be asymptomatics, paucisymptomatics or can decide not to take sick-leave (e.g. for economic reasons). It is important to consider the effect of a small fraction $\varepsilon $ of nodes with high activity $a_S$ which keep their activity when infected. 

We modify the targeted approach considering an \textit{$\varepsilon$-targeted strategy} with $f(a_S)=\theta(a^*-a_S)+\varepsilon \theta(a_S-a^*)$: nodes with $a_S<a^*$ keep $a_I=a_S$ and nodes with $a_S \geq a^*$ have a probability $\varepsilon$ to keep $a_I=a_S$ and $1-\varepsilon$ to zero their infected activity $a_I=0$. In order to make the three strategies comparable, we fix $a^*$ so that in all strategies the fraction of sick-leaving nodes is the same $p$, and we impose $\varepsilon \leq 1-p$ to guarantee $y=a^*/a_m \geq 1$ (see Appendix B). The epidemic threshold is (see Appendix B for the full derivation):
\begin{widetext}
\begin{equation}
r_C^{\varepsilon}= \frac{2 \overline{a_S}}{\overline{a_S} \left(1+\frac{1-y^{1-\nu}+\varepsilon(y^{1-\nu}-\eta^{1-\nu})}{1-\eta^{1-\nu}}\right) + \sqrt{\overline{a_S}^2 \left[ \frac{(y^{1-\nu}-\eta^{1-\nu})(1-\varepsilon)}{1-\eta^{1-\nu}}\right]^2 + 4 \, \overline{a_S^2} \frac{1-y^{2-\nu}+\varepsilon (y^{2-\nu}-\eta^{2-\nu})}{1-\eta^{2-\nu}}}}.
\label{eq:rcE}
\end{equation}
\end{widetext}
The introduction of a small fraction $\varepsilon$ worsens the epidemic threshold compared to the pure-targeted strategy (see Fig. \ref{fig:comparison_2d}(c)). The effect is stronger for higher $\varepsilon$ (see Fig. \ref{fig:Rapp_epsilon}) and for heterogeneous activities ($\nu \sim 1$), while for homogeneous systems the differences are very small. This shows that heterogeneous systems can be controlled effectively by targeted strategies, but they are also susceptible to the presence of a small fraction of uncontrollable nodes. However, even for $\varepsilon>0$ the targeted strategy is still more effective than the uniform one (compare the panels of Fig. \ref{fig:comparison_2d} and Fig. \ref{fig:Rapp_epsilon}), unless $\varepsilon \sim 1-p$.

\begin{figure}
\centering
\includegraphics[width=0.45\textwidth]{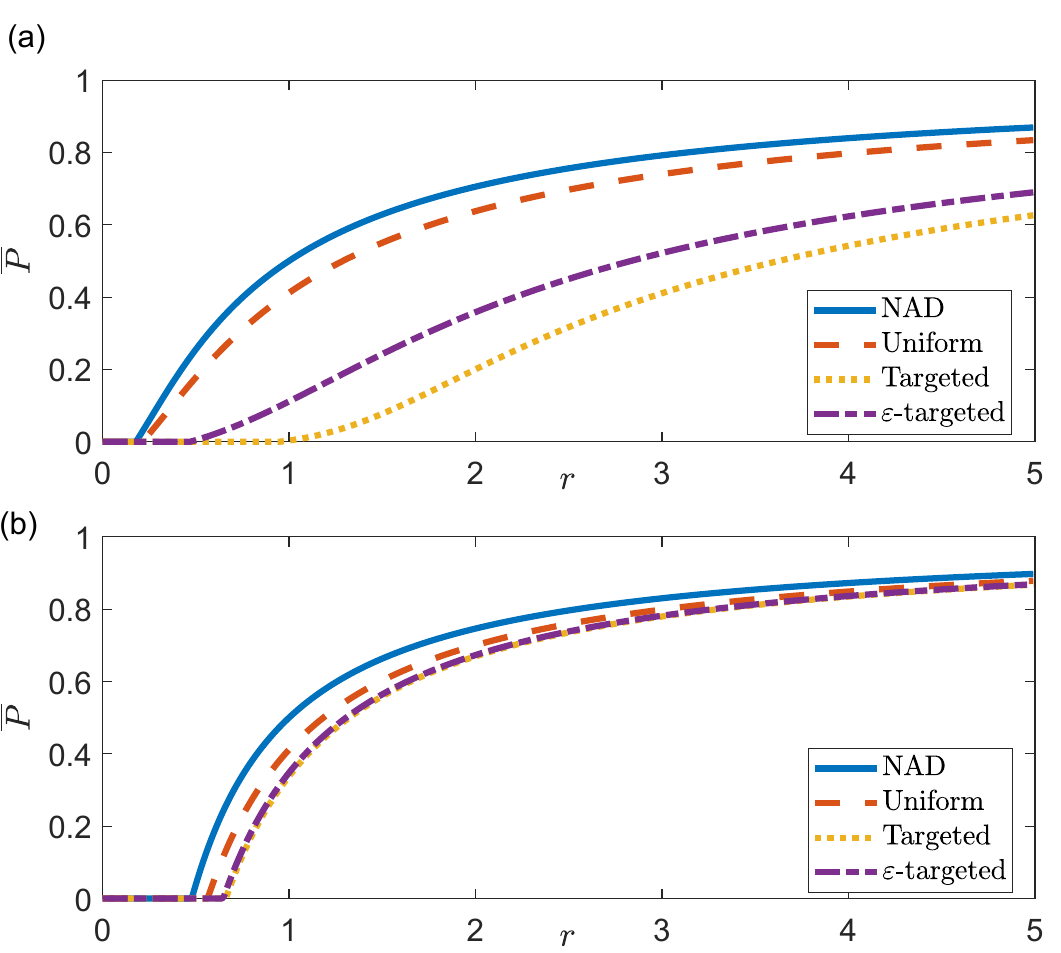}
\caption{\textbf{SIS Epidemic prevalence for sick-leave strategies.} In both panels the SIS epidemic prevalence, $\overline{P}$, is plotted as a function of the control parameter, $r$, for different sick-leave strategies. We fix: the fraction of sick-leaving nodes $p=0.3$, the fraction of high-activity nodes not performing sick-leave $\varepsilon = 0.1$ and the distribution of susceptible activity $\rho_S(a_S) \sim a_S^{-(\nu+1)}$ with lower cut-off $a_m=10^{-3}$, upper cut-off $a_M=1$, $\nu=1$ in panel (a) and $\nu=3.5$ in panel (b). The results are obtained by iterating numerically Eqs. \eqref{eq:prevb}-\eqref{eq:prev}.
}
\label{fig:Prevalence}
\end{figure}

\section{Sick-leave effects on the SIS active phase}\label{sez:sisact}
Often real systems are above the epidemic threshold, with $r>r_C$ (i.e., with $R_0>1$), and it is difficult to move them into the absorbing phase by increasing the epidemic threshold, even with adaptive behaviours. For example, typical values of the basic reproduction number for ILI are $R_0 \in [1,5]$ \cite{biggerstaff2014estimates,cope2018characterising}. It is interesting to investigate the effects of adaptive behaviours and sick-leave strategies on the active phase of the epidemic, i.e. on the epidemic prevalence $\overline{P}$ (average stationary infection probability).

In Fig. \ref{fig:Prevalence} we plot the prevalence, calculated by iterating numerically Eqs. \eqref{eq:prevb}-\eqref{eq:prev} (see Appendix A), as a function of $r$, for fixed $\nu$ and $p=0.3$, which represents a realistic estimation for ILI epidemics \cite{ariza2018healthcareseeking,eames2010impact, vankerckhove2013impact}. All the sick-leave strategies lower the epidemic prevalence compared to the NAD case: however, for heterogeneous activities ($\nu=1$) the three approaches produce significantly different results (Fig. \ref{fig:Prevalence}(a)) with the targeted approach much more effective, followed by the $\varepsilon$-targeted one and then by the uniform one (worst case). On the contrary for homogeneous activities ($\nu=3.5$), the strategies are almost equivalent and ineffective both in increasing the threshold $r_C$ and in lowering the epidemic prevalence $\overline{P}$, since the curves overlap (Fig. \ref{fig:Prevalence}(b)). 

\section{SIR model on adaptive activity-driven networks}\label{sez:SIR_r}
Infectious diseases can produce immunity in infected individuals, in this case the epidemic is better described by the SIR epidemic model. In the SIR version of the adaptive activity-driven network, each node $i$ is characterized by three parameters $(a_S^i,a_I^i,a_R^i)$, respectively the activity of node $i$ in the susceptible $S$, infected $I$ and recovered state $R$, and they are drawn from the joint distribution $\rho_{SIR}(a_S,a_I,a_R)$. At first the nodes are disconnected, then the network evolves with a continuous dynamics of link formation: each node activates with a Poissonian dynamics, with rate $a_S$ if the node is susceptible, with rate $a_I$ if infected and with rate $a_R$ if recovered. An active node creates one link with a randomly chosen node: if the nodes involved in the contact are one infected $I$ and one susceptible $S$, the susceptible node gets infected with probability $\lambda$, $S+I \xrightarrow[]{\lambda} 2I$, otherwise nothing happens during the interaction. Then the contact is removed. Infected nodes recover with rate $\mu$, through a Poissonian spontaneous recovery process, $I \xrightarrow[]{\mu} R$, and they get immunity (i.e. they are no longer infectious and cannot get anymore infected). Hereafter we will assume that the recovered nodes regain the activity they had before the infection $a_R^i=a_S^i$: $\rho_{SIR}(a_S,a_I,a_R)=\rho(a_S,a_I) \delta(a_R-a_S)$, where $\delta( \cdot )$ indicates the Dirac delta function. This choice has no significant implication, as far as the epidemic dynamics is concerned: recovered nodes no longer enter into the contagion process. However, this choice significantly influences the average activity level of the population and the network evolution (see Section \ref{sez:rel}). 

\begin{figure}
\centering
\includegraphics[width=0.45\textwidth]{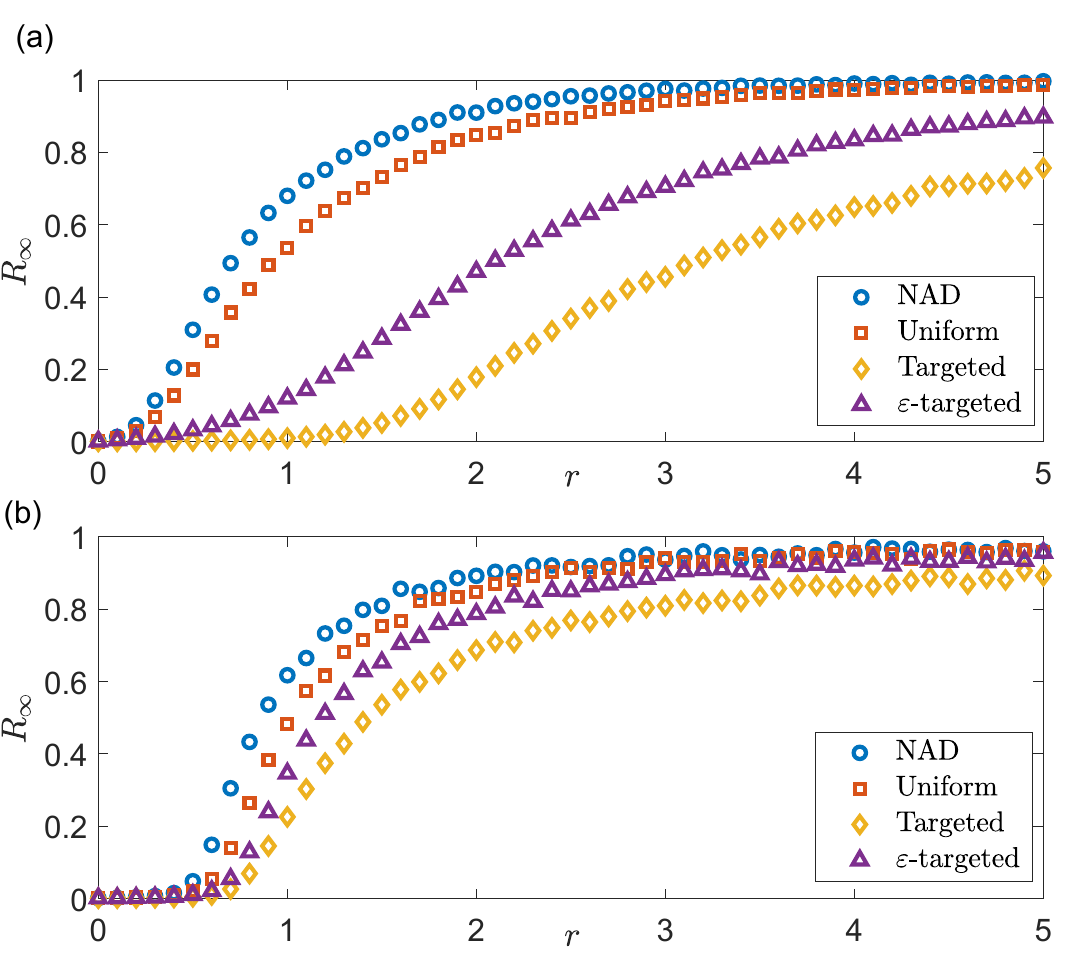}
\caption{\textbf{SIR Epidemic final-size for sick-leave strategies.} 
In both panels the SIR epidemic final-size, $R_{\infty}$, is plotted as a function of the control parameter, $r$, for different sick-leave strategies. We fix: the fraction of sick-leaving nodes $p=0.3$, the fraction of high-activity nodes not performing sick-leave $\varepsilon = 0.1$ and the distribution of susceptible activity $\rho_S(a_S) \sim a_S^{-(\nu+1)}$ with lower cut-off $a_m=10^{-3}$, upper cut-off $a_M=1$, $\nu=1$ in panel (a) and $\nu=3.5$ in panel (b). The results are obtained through numerical simulations over networks of $N=10^3$ nodes: each point is obtained by averaging over at least $10^3$ realizations of the dynamical evolution and of the underlying network, until the errors on $R_{\infty}$ and on the maximum of the infection peak, $\overline{P}_{max}$, are both lower than 1\%.
}
\label{fig:PrevalenceSIR}
\end{figure}

The epidemic threshold $r_C$ of the SIR and SIS models is the same, because of the completely mean-field nature of the model \cite{tizzani2018memory,Valdano2015}. On the contrary, the active phase is different: the SIR model lacks a stationary endemic state and we cannot obtain information on the active phase of the epidemic in an analytic way, thus we perform extensive numerical simulations. The effects of the sick-leave strategies on the SIR active phase can be estimated by several quantities: (i) the epidemic final size $R_{\infty}$, i.e. the total fraction of nodes infected throughout the epidemic, which is the order parameter of the phase transition; (ii) the temporal evolution of the average infection probability $\overline{P}(t)$, in particular (iii) the height of the infection peak $\overline{P}_{max}$, (iv) the time $t_{max}^I$ at which the peak occurs and (v) the width of the peak are relevant for the impact of the epidemic on the healthcare system and for planning countermeasures \cite{anderson2020individual}. Note that in the SIR model the epidemic final size, $R_{\infty}$, coincides with the final fraction of nodes recovered, since susceptible nodes can get infected only once in the epidemic process and after the infection they all recover by gaining immunity.

We perform numerical simulations of the SIR process on the adaptive activity-driven network. We assign to each of the $N$ nodes, their activities $a_S$ and $a_I$ according to the sick-leave strategy adopted: the network dynamics and the epidemic spreading are simulated by using a Gillespie-like algorithm \cite{gillespie1976general,mancastroppa2019burstiness,mancastroppa2020active}. We first let the network evolve without epidemics so that network dynamics is relaxed to the equilibrium. Then the epidemics is initialized, by infecting the node with the highest activity $a_I$ \cite{nature2013boguna}. See Appendix C for a detailed sketch of the numerical implementation. The results are averaged over several realizations of the dynamical evolution and of the underlying network.\\

\begin{figure}
\centering
\includegraphics[width=0.45\textwidth]{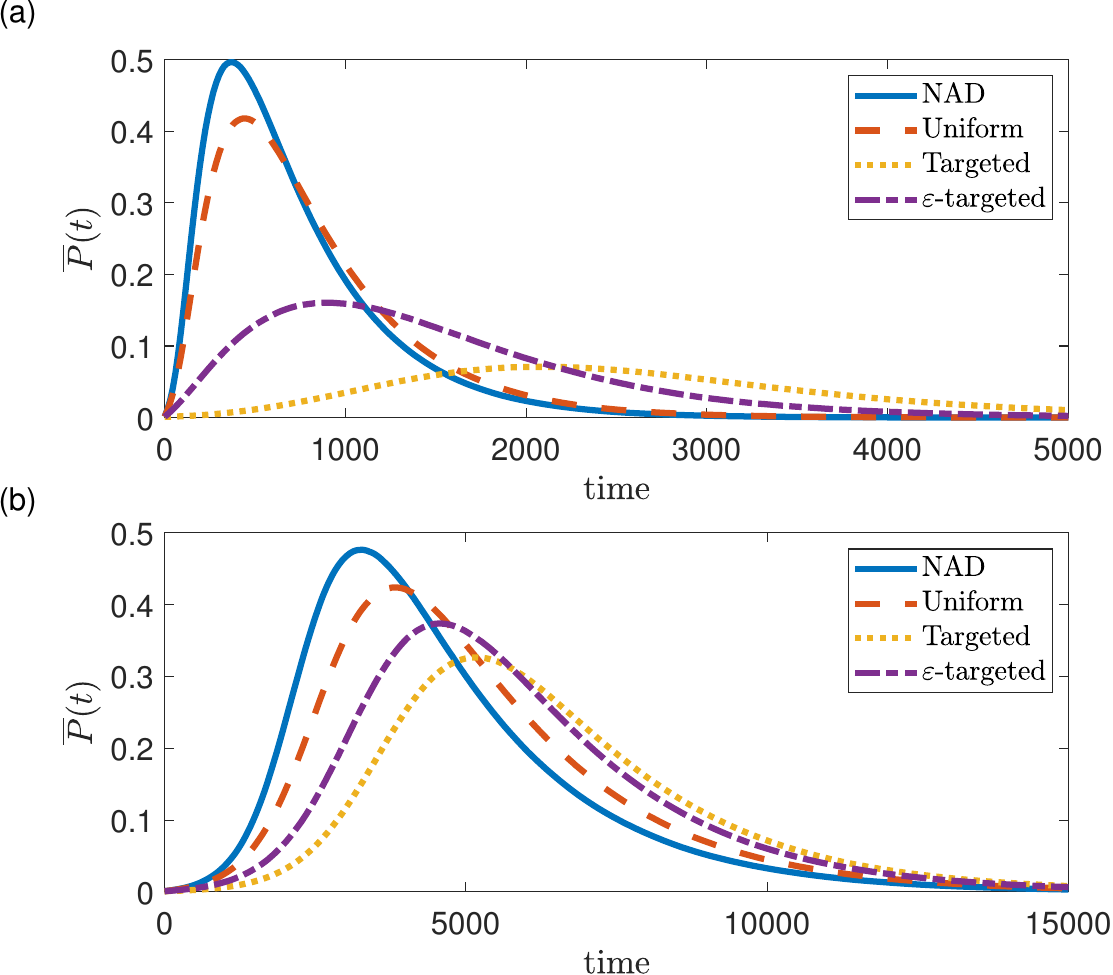}
\caption{\textbf{Infection peak for sick-leave strategies.} In both panels we plot the fraction of infected nodes, $\overline{P}(t)$, as a function of time for several sick-leave strategies. We fix: the control parameter $r=3$, the fraction of sick-leaving nodes $p=0.3$, the fraction of high-activity nodes not performing sick-leave $\varepsilon = 0.1$ and the distribution of susceptible activity $\rho_S(a_S) \sim a_S^{-(\nu+1)}$ with lower cut-off $a_m=10^{-3}$, upper cut-off $a_M=1$, $\nu=1$ in panel (a) and $\nu=3.5$ in panel (b). The results are obtained through numerical simulations over networks of $N=10^3$ nodes: each curve is obtained by averaging over at least $10^3$ realizations of the dynamical evolution and of the underlying network, until the errors on the epidemic final-size, $R_{\infty}$, and on the maximum of the infection peak, $\overline{P}_{max}$, are both lower than 1\%.
}
\label{fig:comparison_temporal}
\end{figure}

\begin{figure*}
\centering
\includegraphics[width=0.95\textwidth]{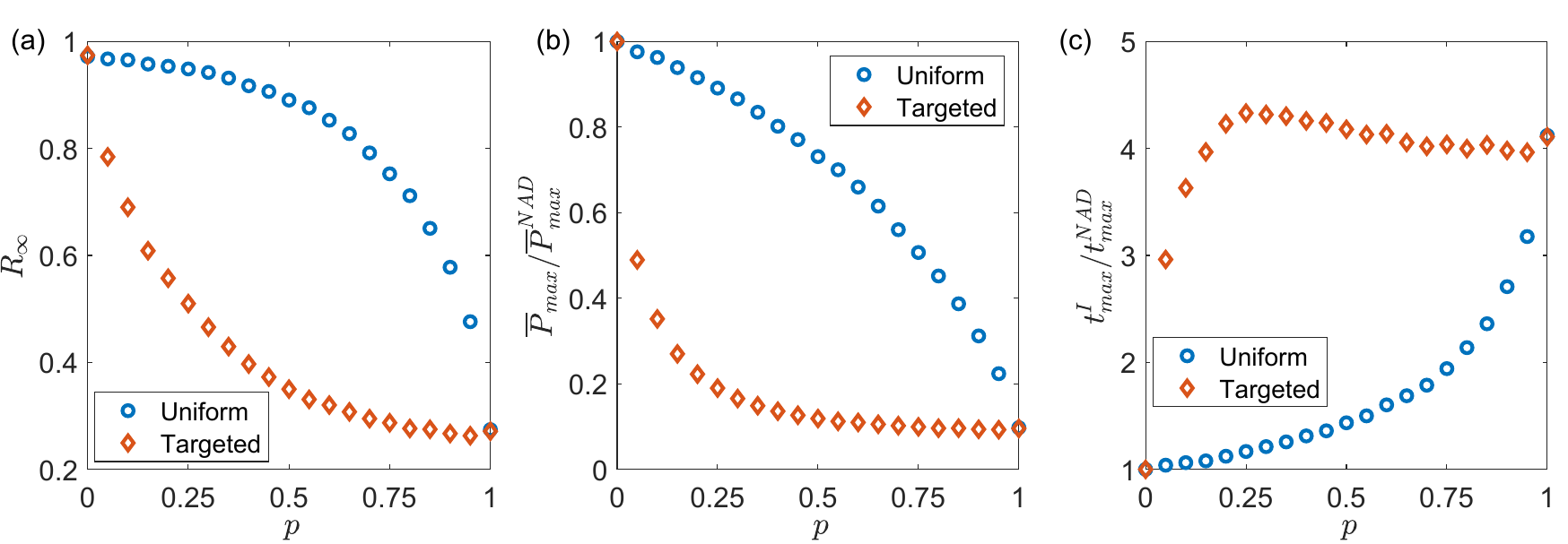}
\caption{\textbf{Effects of sick-leave strategies intensity on the epidemic active phase.} In panel (a) we plot the epidemic final-size, $R_{\infty}$, as a function of the fraction of sick-leaving nodes $p$; in panel (b) the ratio $\overline{P}_{max}/\overline{P}_{max}^{NAD}$ between the infection peak height in the adaptive and in the non-adaptive (NAD) case, is plotted as a function of $p$; in panel (c) we plot the ratio $t_{max}^I/t_{max}^{NAD}$ between the infection peak time in the adaptive and in the NAD case, as a function of $p$. In all panels are plotted both the uniform and targeted case, we fix the control parameter $r=3$ and the distribution of susceptible activity $\rho_S(a_S) \sim a_S^{-(\nu+1)}$ with lower cut-off $a_m=10^{-3}$, upper cut-off $a_M=1$ and $\nu=1$. The results are obtained through numerical simulations over networks of $N=10^3$ nodes; each point of the adaptive and NAD case is obtained by averaging over at least $10^3$ realizations of the dynamical evolution and of the underlying network, until the errors on $R_{\infty}$ and on $\overline{P}_{max}$ are both lower than 1\%.
}
\label{fig:ratios_nu1}
\end{figure*}

In Fig. \ref{fig:PrevalenceSIR} we plot $R_{\infty}$ as a function of $r$ for each sick-leave strategy, fixing $p=0.3$ and $\varepsilon=0.1$. All the sick-leave strategies generally lower the epidemic final-size, reducing the number of people been infected during the epidemic. However, for $\nu = 1$ the three strategies performances are significantly different (Fig. \ref{fig:PrevalenceSIR}(a)): the targeted approach is the most effective one, followed by the $\varepsilon$-targeted strategy and finally by the uniform approach (worst case). On the contrary for homogeneous activities $\nu=3.5$ (Fig. \ref{fig:PrevalenceSIR}(b)), the strategies are almost equivalent and poorly effective, both in increasing the threshold $r_C$ and in decreasing the epidemic final-size. 

In panel (a) of Fig. \ref{fig:comparison_temporal} we plot the fraction of infected nodes over time $\overline{P}(t)$ fixing $r=3$ (as for realistic ILI \cite{biggerstaff2014estimates,cope2018characterising}), $p=0.3$ and $\nu=1$. The targeted strategy is extremely effective in delaying and reducing the impact of the epidemic, flattening the infection peak, on the contrary the uniform strategy slightly affects the infection peak. These strong differences between strategies hold as long as the activities are heterogeneous: see Fig. \ref{fig:comparison_temporal}(b) for a comparison with the homogeneous case at $\nu=3.5$, where the differences are significantly reduced. \\

In order to study the effect of the strategy intensity, $p$, we focus on a realistic system with heterogeneous activities $\nu=1$ \cite{ubaldi2016asymptotic,perra2012activity,ribeiro2013quantifying, karsai2014time} in the active phase $r=3$, as for realistic ILI \cite{biggerstaff2014estimates,cope2018characterising}. In Fig. \ref{fig:ratios_nu1} we plot $R_{\infty}$, $\overline{P}_{max}/\overline{P}_{max}^{NAD}$ (ratio between the infection peak height in the adaptive and NAD cases) and $t_{max}^I/t_{max}^{NAD}$ (ratio between the temporal occurrence of the infection peak in the adaptive and NAD cases) as a function of $p$. Both uniform and targeted strategies at most can reduce the epidemic final-size of about 75\% and the infection peak of about 90\%, while they can delay the peak of $\sim 4.5$ times. However, only the targeted strategy is effective at small $p$ allowing to intervene with mild measures: when realistically 25\% of the population performs sick-leave \cite{ariza2018healthcareseeking}, the time of infection peak occurrence is more than quadrupled, the epidemic final-size is reduced to $50\%$ of the population and the infection peak height is reduced by $80\%$. On the contrary, the uniform strategy is extremely ineffective: in order to obtain the same results, at least 95\% of the population is needed to perform sick-leave.

\section{Effects of sick-leave on population activity and absenteeism levels}\label{sez:rel}

When epidemic control measures are implemented, it is important to estimate whether and to what extent these compromise the operativeness of the population \cite{Massaro2018,aguilar2021absenteism_strategies,hollingsworth2011mitigation,fenichel2013economic,bonaccorsi2020,Thommes2016,gianino2017,Ip2015}. We now study the effects of the sick-leave strategies on the average population activity level $\langle a \rangle (t) = (\sum_{i \in I(t)} a_I^i + \sum_{j \in S(t),R(t)} a_S^j)/N$, on the fraction of simultaneously sick-leaving nodes $A(t)$, and on their evolution. These are the relevant quantities to determine if the sick-leave strategies compromise the operativeness and activity of the system \cite{hollingsworth2011mitigation,fenichel2013economic,Ip2015,gianino2017,Thommes2016}. \\

\begin{figure*}
\centering
\includegraphics[width=\textwidth]{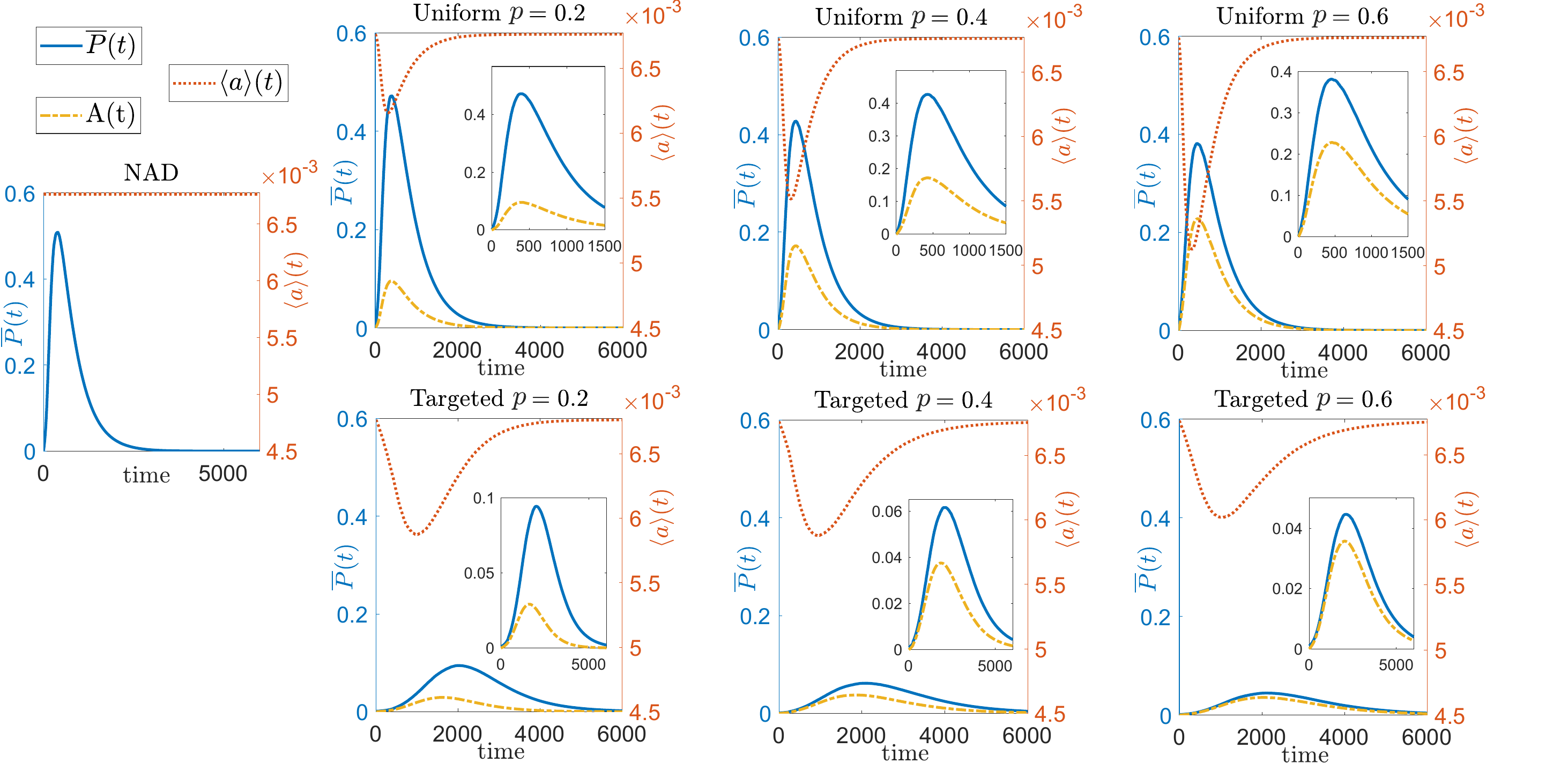}
\caption{\textbf{Effects of sick-leave strategies on the population activity and absenteeism dynamics.} In each panel it is plotted the fraction of infected individuals, $\overline{P}(t)$ (blue solid line - left y-axis), the fraction of absent nodes, $A(t)$ (yellow dash-dotted line - left y-axis), and the average activity, $\langle a \rangle (t)$ (red dotted line - right y-axis), as a function of time $t$. The first panel on the left side correspond to the non-adaptive (NAD) case, while the panels on the right side correspond to the uniform (first row) and targeted (second row) strategies for several values of the fraction of sick-leaving nodes $p$ (see above each panel). In each panel, the inset is a zoom on the infection and absenteeism peaks. We fix the control parameter $r=3$  and the distribution of susceptible activity $\rho_S(a_S) \sim a_S^{-(\nu+1)}$ with lower cut-off $a_m=10^{-3}$, upper cut-off $a_M=1$ and $\nu=1$. The results are obtained through numerical simulations over networks of $N=10^3$ nodes: each curve is obtained by averaging over at least $10^3$ realizations of the dynamical evolution and of the underlying network, until the errors on the epidemic final-size, $R_{\infty}$, and on the maximum of the infection peak, $\overline{P}_{max}$, are both lower than 1\%.
}
\label{fig:temporal_nu1}
\end{figure*}

\begin{figure*}
\centering
\includegraphics[width=0.6\textwidth]{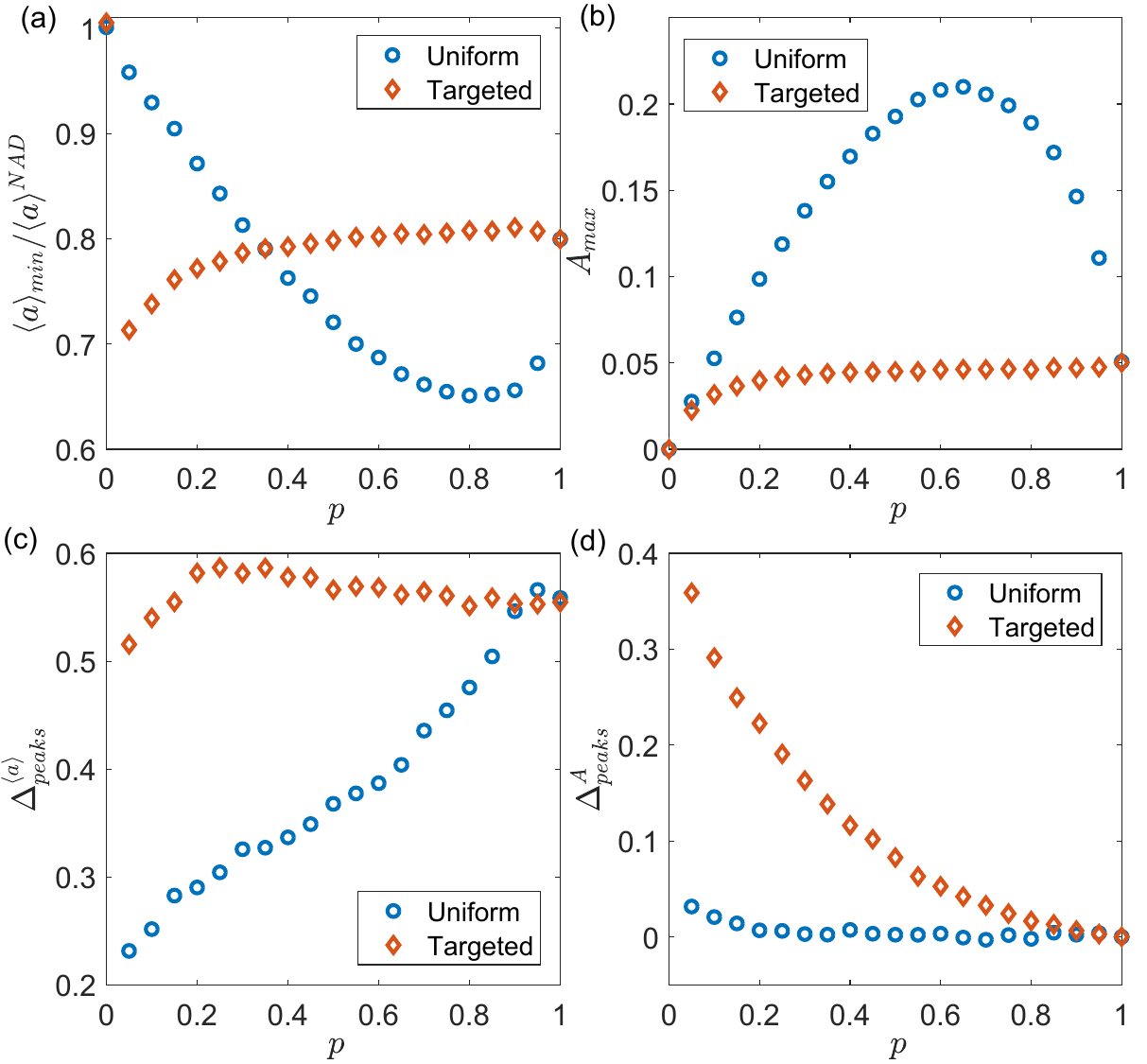}
\caption{\textbf{Effects of sick-leave strategies on the population activity and absenteeism levels.} In panel (a) we plot the ratio $\langle a \rangle_{min}/\langle a \rangle^{NAD}$ between the minimum average activity in the adaptive and in the NAD case, as a function of the fraction of sick-leaving nodes, $p$; in panel (b) we plot the maximum fraction of simultaneously absent nodes $A_{max}$ as a function of $p$; in panel (c) it is plotted the $\Delta^{\langle a \rangle}_{peaks}=(t_{max}^I-t_{min}^{\langle a \rangle})/t_{max}^I$ relative time delay between the minimum in the population activity $t_{min}^{\langle a \rangle}$ and the infection peak $t_{max}^I$ as a function of $p$; in panel (d) it is plotted the $\Delta^A_{peaks}=(t_{max}^I-t_{max}^A)/t_{max}^I$ relative time delay between the maximum in the absenteeism $t_{max}^A$ and the infection peak $t_{max}^I$ as a function of $p$. In all panels are plotted both the uniform and targeted case, we fix the control parameter $r=3$ and the distribution of susceptible activity $\rho_S(a_S) \sim a_S^{-(\nu+1)}$ with lower cut-off $a_m=10^{-3}$, upper cut-off $a_M=1$ and $\nu=1$. The results are obtained through numerical simulations over networks of $N=10^3$ nodes; each point of the adaptive and NAD case is obtained by averaging over at least $10^3$ realizations of the dynamical evolution and of the underlying network, until the errors on $R_{\infty}$ and on $\overline{P}_{max}$ are both lower than 1\%.
}
\label{fig:ratios_nu2}
\end{figure*}

In Fig. \ref{fig:temporal_nu1} we plot the temporal evolution of the infection peak $\overline{P}(t)$, the fraction of absent node $A(t)$ and the population activity level $\langle a \rangle (t)$. In the NAD case the activity level is constant and there is no absenteeism curve, while both in the uniform and targeted strategies, $\langle a \rangle(t)$ and $A(t)$ features respectively a minimum and a maximum. Interestingly, the minimum of activity anticipate the infection peak, especially in the targeted case. The first nodes to be infected are, indeed, the most active ones which are removed by the sick-leave mechanism, so that the activity immediately collapses in the early stages. Then when nodes of higher activity start to recover, also the other nodes are infected, generating the infection peak, which thus is delayed compared to the minimum in activity. Hence, the temporal shift is relevant only in heterogeneous networks and is more pronounced in the case of a targeted strategy, since all the most active nodes perform sick-leave. In the targeted strategy an anticipation is also observed in the absenteeism peak compared to the infection one (Fig. \ref{fig:temporal_nu1}). This time shift is crucial for example in hospitals or schools, since the population returns almost completely operative during the infection peak, thus providing essential services in the most critical phase.

The observed time shift between human activity and the infection peak was investigated in different settings: for example, several studies of absenteeism in schools and workplaces show signs of a delay \cite{bollaerts2010timeliness,besculides2005evaluation,Donaldson2021}, while other studies do not detect this delay \cite{graitcer2012effects}. Moreover, a temporal shift in the peak of infection between different age groups is often observed: children experience an early infection peak compared to adults, due to their high mixing rate and their high activity (i.e. different activity classes in our model) \cite{mossong2008social,peters2014relative,apolloni2013age}.

Fig. \ref{fig:ratios_nu2}(a) shows that both uniform and targeted strategies at most produce a reduction of 30-35\% in the average activity level despite the intensity of the strategies adopted ($p$). 
In the targeted strategy the activity after a sudden drop at $p \sim 0$ starts to recover and reaches a saturation at $p \sim 0.3$, where the number of infected nodes is significantly reduced (see Fig. \ref{fig:ratios_nu1}(a)-(b)). Therefore, the targeted strategy with $p \sim 0.3$, which is consistent with its realistic value \cite{ariza2018healthcareseeking}, produce an important reduction of epidemic spreading without reducing excessively the activity of the system (reduction of about 20\% of the average activity level). On the contrary, the uniform strategy produces a similar level of reduction in the average activity for $p \sim 0.3$ without any significant reduction in the epidemic transmission.

In both the uniform and targeted strategies the activity minimum anticipates the infection peak, since the most active nodes are usually the first to get infected. However, in the uniform strategy the relative time delay $\Delta_{peaks}^{\langle a \rangle}=(t_{max}^I-t_{min}^{\langle a \rangle})/t_{max}^I$ between the activity minimum and the infection peak increases with $p$ reaching $0.6$ (see Fig. \ref{fig:ratios_nu2}(c)). Instead, in the targeted case we get a large delay, about $0.6$, independently of $p$, since high activity nodes always take sick-leave also for small $p$: even for low intensities of the mitigation measure, the infection peak is distinct from the minimum in population activity, unlike the uniform strategy. \\

So far we have considered the impact of sick-leave on the average activity of the population, however in real systems it is complicated to estimate the reduction in the system activity, as it would require knowing the social activity of each sick-leaving node. On the contrary, one typically has easy access to the level of absenteeism over time in a specific setting \cite{bollaerts2010timeliness,besculides2005evaluation,Donaldson2021,graitcer2012effects}. We now focus on the number $A(t)$ of individuals simultaneously absent in time and its maximum $A_{max}$ (see Fig. \ref{fig:temporal_nu1}).

The targeted strategy keeps very low the number of simultaneously absent nodes ($3\%$-$5\%$) also at $p \sim 0.3$ (Fig. \ref{fig:ratios_nu2}(b)) where significant containment is obtained. On the contrary, the uniform strategy produce a larger level of absenteeism ($10\%$-$20\%$), especially for realistic $p$ values and for $p$ values that produce a sufficient epidemic mitigation for this strategy.

Furthermore, in the uniform strategy the relative time delay $\Delta_{peaks}^A=(t_{max}^I-t_{max}^A)/t_{max}^I$ between the peak of infection and the maximum of absenteeism is approximately zero for every value of $p$ (see Fig. \ref{fig:ratios_nu2}(d)), hence during the most critical phase of the epidemic the population is always characterized by high levels of absenteeism. On the contrary, the targeted strategy produces an anticipation of the absenteeism peak on the infection peak, with a relative delay of approximately $\Delta_{peaks}^A \sim 0.2$ for realistic $p \sim 0.3$, which then decreases to zero when increasing $p$. \\

In realistic heterogeneous populations sick-leave strategies targeted on the most at-risk nodes are significantly more effective compared to uniform strategies, especially for mild control. In particular, for a fraction $p \simeq 0.3$ of sick-leaving nodes, the targeted strategy strongly weakens the epidemic, while keeping low the absenteeism level and high the population activity level. Moreover, it increases the time shifts in the minimum of population activity and in the maximum of absenteeism, with respect to the infection peak, suggesting that the population returns almost completely operative during the infection peak. Therefore, $p \simeq 0.3$ can be considered an effective fraction for a targeted sick-leave prescription. On the contrary, uniform strategies are ineffective in facing the epidemic and also produce a strong deterioration in population operativeness, with high level of absenteeism and low system activity, especially during the infection peak. 

\section{Conclusions}
This work provides a framework for understanding the interplay between social interaction dynamics, epidemic spreading and adaptive behaviours of individuals, allowing us to investigate the effects of adaptation and of mild-to-moderate control measures, such as \textit{sick-leave}. The combined effects of these interventions on the epidemics and on the interaction dynamics are crucial for reaching the control of epidemics \cite{anderson2020individual,fraser2004,mancastroppa2020active} while preserving the essential services and systems, i.e. the population overall operativeness and activity \cite{hollingsworth2011mitigation,fenichel2013economic}.

Through adaptive temporal networks, we formulate a general framework which models a wide range of different adaptive behaviours and mitigation strategies, observed and implemented in real populations exposed to epidemics. We derive an analytical estimate of the epidemic threshold for arbitrary adaptive behaviours and we highlight the crucial role of correlations between individuals behaviour in the infected and in the susceptible state. We focus on the sick-leave practice and we compare the effects of several strategies on the SIS and SIR epidemic processes, showing the critical relevance of heterogeneity: in homogeneous systems, targeted and uniform strategies are equivalent and both are poorly effective; on the contrary, in heterogeneous networks, targeted strategies (over most at-risk nodes) are considerably more effective than non-targeted ones, especially for small fractions of sick-leaving nodes, that is for mild control measures. These results are robust to even small errors in detecting high-risk nodes in the targeted strategy. Moreover, targeted strategies are both effective in flattening the infection peak and delaying it, also increasing the time difference with the maximum of absenteeism and the minimum in the activity of the population, so that the population returns almost completely operative during the infection peak. The activity of the system is not excessively reduced, guaranteeing the operativeness of the system, and the strategy keeps the number of simultaneously absent nodes low. On the contrary, the uniform strategy requires very high control, producing high levels of absenteeism and a strong deterioration in the activity of the population, especially during the infection peak, i.e. the most critical phase of the epidemic.

Our results, despite the simplicity of the model, provide crucial insights on the implementation of mitigation strategies and adaptive behaviours, and on their impact on population operativeness, opening new perspectives in the control of epidemic spreading on realistic adaptive temporal networks. The model can be modified to account for other realistic features, such as a differentiated adaptive behaviour for strong and weak ties \cite{tizzani2018memory,karsai2014time,Sun2015cont}, as detected in several real populations \cite{ariza2018healthcareseeking,karsai2014time}, heterogeneous temporal patterns \cite{mancastroppa2019burstiness,ubaldi2017burstiness}, or additional mild interventions  \cite{mancastroppa2020active,Mancastroppa2021CT,Mancastroppa2022CT}. Moreover, the introduced model paves the way for the modelling of even more complicated behaviours that can be easily implemented through the proposed framework, for example the introduction of multiple groups that have intermediate activity reduction when infected \cite{eames2010impact,vankerckhove2013impact,ariza2018healthcareseeking}, or the modelling of groups of people who chase the infection and increase their social activity when infected \cite{swine2009,covid_2,pox}.

\section{Acknowledgements}
This research was granted by University of Parma through the action \textit{Bando di Ateneo 2022 per la ricerca} co-funded by MUR-Italian Ministry of Universities and Research - D.M. 737/2021 - PNR - PNRR - NextGenerationEU (project \textit{'Collective and self-organised dynamics: kinetic and network approaches'}). M.M. and V.C. acknowledge support from the Agence Nationale de la Recherche (ANR) project DATAREDUX (ANR-19-CE46-0008).

\section*{Appendix A: Analytical derivation of the general epidemic threshold and epidemic prevalence}
\label{app:A}
In this Appendix we provide the detailed analytical derivation of the asymptotic epidemic prevalence and of the epidemic threshold of the SIS epidemic model on the adaptive activity-driven network described in Section \ref{sez:model}. 

Each node is assigned with two parameters $(a_S,a_I)$ drawn from the general joint distribution $\rho(a_S,a_I)$. We apply an activity-based mean-field approach (ABMF), which is exact since correlations are destroyed over time by link reshuffling. We divide the population into classes with the same activities $(a_S,a_I)$ and consider the probability $P_{a_S,a_I}(t)$ that a node in class $(a_S,a_I)$ is infected at time $t$. In the thermodynamic limit, the epidemic dynamics is governed by Eq. \eqref{eq:dynamics}. Focusing on the SIS asymptotic steady state and on the asymptotic epidemic prevalence $P_{a_S,a_I}^0=\lim\limits_{t \to \infty} P_{a_S,a_I}(t)$, the Eq. \eqref{eq:dynamics} becomes:
\begin{equation}
\partial_t P_{a_S,a_I}^0=-\mu P_{a_S,a_I}^0 + \lambda \left(1-P_{a_S,a_I}^0 \right)\left[ a_S \overline{P} + \overline{a_I P}  \right],
\label{eq:dynamics2}
\end{equation}
where $\overline{P}=\int da_S \int da_I \rho(a_S,a_I) P_{a_S,a_I}^0$ is the average asymptotic epidemic prevalence and $\overline{a_I P}=\int da_S \int da_I \rho(a_S,a_I) a_I P_{a_S,a_I}^0$ is the average asymptotic infectivity. The first term on the right-hand side corresponds to recovery processes, while the second term corresponds to infection processes, according to the two possible paths: a susceptible node of class $(a_S,a_I)$ activates and contacts an infected node of any class $(a_S',a_I')$ or an infected node of any $(a_S',a_I')$ class activates and contacts a susceptible node of the $(a_S,a_I)$ class. By multiplying both sides of Eq. \eqref{eq:dynamics2} by $\rho(a_S,a_I)$ and integrating over $(a_S,a_I)$, we obtain the equation for $\overline{P}$:
\begin{equation}
\partial_t \overline{P} = \left[ -\mu + \lambda \left( \overline {a_S} - \overline {a_S P} \right) \right] \overline{P} + \lambda \left(1- \overline{P} \right) \overline{a_I P},
\label{eq:P}
\end{equation}
where $\overline{g(a_S,a_I) P} = \int da_S \int d a_I \rho(a_S,a_I) P_{a_S,a_I}^0 g(a_S,a_I)$ and $\overline{g(a_S,a_I)} = \int da_S \int da_I \rho(a_S,a_I) g(a_S,a_I)$. Similarly, by multiplying both sides of Eq. \eqref{eq:dynamics2} by $a_I \rho(a_S,a_I)$ and integrating over $(a_S,a_I)$, we obtain the equation for $\overline{a_I P}$:
\begin{equation}
\partial_t \overline{a_I P} = \lambda \left( \overline {a_I a_S} - \overline {a_I a_S P} \right) \overline{P} + \left[ -\mu + \lambda \left(\overline{a_I}- \overline{a_I P} \right) \right] \overline{a_I P}.
\label{eq:aP}
\end{equation}
Eqs. \eqref{eq:P}-\eqref{eq:aP} compose a set of non-linear differential equations which admits the absorbing state $(\overline{P},\overline{a_I P})=(0,0)$ as a stationary solution. We apply a linear stability analysis around the absorbing state to obtain the epidemic threshold $\lambda_C$. We linearize the equations around the absorbing state and obtain:
\begin{equation}
\partial_t \overline{P} =\left[ -\mu + \lambda \overline {a_S} \right] \overline{P} + \lambda \overline{a_I P},
\label{eq:sys1}
\end{equation}
\begin{equation}
\partial_t \overline{a_I P} = \left[ -\mu + \lambda \overline{a_I} \right] \overline{a_I P} + \lambda \overline {a_I a_S} \overline{P}.
\label{eq:sys2}
\end{equation}
This system has Jacobian matrix $J$:
\begin{equation}
J=
\begin{bmatrix}
-\mu+\lambda \overline{a_S} & \lambda\\
\lambda \overline{a_I a_S} & -\mu + \lambda \overline{a_I}
\end{bmatrix},
\end{equation}
which admits eigenvalues:
\begin{equation}
\xi_{1,2}=\frac{1}{2} \left[ \lambda(\overline{a_I}+\overline{a_S})-2 \mu \pm \lambda \sqrt{(\overline{a_S}-\overline{a_I})^2+4 \overline{a_I a_S}} \right].
\end{equation}
The stability of the absorbing state is obtained by imposing that the maximum eigenvalue is negative $\xi_{max}<0$. This condition produces the following epidemic threshold:
\begin{equation}
\lambda_C= \frac{2 \mu}{\overline{a_S} + \overline{a_I} + \sqrt{(\overline{a_S} - \overline{a_I})^2 + 4 \, \overline{a_I a_S}}}.
\label{eq:thr2}
\end{equation}
This threshold is the same as in Eq. \eqref{eq:thr}, which is formulated in terms of $r=\lambda \overline{a_S}/\mu$: it is exact since the model is exactly mean-field, and it holds for arbitrary $\rho(a_S,a_I)$. For example, the non-adaptive (NAD) case corresponds to the case where no adaptive behaviours nor response measures are implemented in the population, i.e. $\rho(a_S,a_I)= \rho_S(a_S) \delta(a_I-a_S)$, where $\delta( \cdot )$ is the Dirac delta function, thus all nodes feature $a_S=a_I$. In this case $\overline{a_I}=\overline{a_S}$ and $\overline{a_I a_S}=\overline{a_S^2}$: thus, substituting in Eq. \eqref{eq:thr2}, the epidemic threshold is:
\begin{equation}
\lambda_C^{NAD}=\frac{\mu}{\overline{a_S} + \sqrt{\overline{a_S^2}}}.
\end{equation}
as in Eq. \eqref{eq:NAD} and as obtained in \cite{perra2012activity}.

Eqs. \eqref{eq:dynamics2}-\eqref{eq:P} can provide analytically the epidemic prevalence in the steady state. By using the definition of steady state, $\partial_t P_{a_S,a_I}^0=0$ and $\partial_t \overline{P}=0$, and replacing this in Eqs. \eqref{eq:dynamics2}-\eqref{eq:P} we obtain the explicit equations for $P_{a_S,a_I}^0$ and $\overline{P}$ as shown in Eqs. \eqref{eq:prevb}-\eqref{eq:prev}. The Eq. \eqref{eq:prev} for $\overline{P}$ depends on $\overline{a_I P}$ and on $\overline{a_S P}$, however it is not possible to close these equations analytically since each variable $\overline{a_I^n P}$ $\forall n$ depends on $\overline{a_I^n a_S P}$ and each variable $\overline{a_S^n P}$ $\forall n$ depends on $\overline{a_S^{n+1} P}$, constituting a set of infinite coupled equations. Therefore, we obtain $\overline{P}$ by iterating numerically Eqs. \eqref{eq:prevb}-\eqref{eq:prev}.

\section*{Appendix B: Analytical derivation of the epidemic threshold for the sick-leave strategies}
\label{app:B}
In this Appendix we provide in details the analytical derivation of the epidemic threshold for the different sick-leave strategies. Hereafter, we will indicate with $\delta( \cdot )$ the Dirac delta function.

Initially we consider the general sick-leave practice, presented in Section \ref{sez:sick}: in this case $\rho(a_S,a_I)=\rho_S(a_S) \rho_{I|S}(a_I|a_S)$ with $\rho_S(a_S) \sim a_S^{-(\nu+1)}$, where $a_S \in [a_m,\eta a_m]$ with $\eta \in (1,\infty)$, and $\rho_{I|S}(a_I|a_S)= f(a_S) \delta(a_I-a_S) + (1-f(a_S)) \delta(a_I)$. In this case:
\begin{equation}
\begin{aligned}
\overline{a_I}& = \int da_S \, da_I \rho_S(a_S) a_I f(a_S) \delta(a_I-a_S) \\ & + \int da_S \, da_I \rho_S(a_S) a_I (1-f(a_S)) \delta(a_I) = \overline{a_S f(a_S)},
\end{aligned}
\label{eq:aI_sick}
\end{equation}
\begin{equation}
\begin{aligned}
\overline{a_I a_S}&= \int da_S \, da_I \rho_S(a_S) a_I a_S f(a_S) \delta(a_I-a_S) \\ 
&+ \int da_S \, da_I \rho_S(a_S) a_I a_S (1-f(a_S)) \delta(a_I) = \overline{a_S^2 f(a_S)}.
\end{aligned}
\label{eq:aSaI_sick}
\end{equation}
Therefore, by replacing Eqs. \eqref{eq:aI_sick}-\eqref{eq:aSaI_sick} into Eq. \eqref{eq:thr2}, we obtain:
\begin{equation}
\lambda_C=\frac{2 \mu}{\overline{a_S} + \overline{a_S f(a_S)} + \sqrt{\left(\overline{a_S} - \overline{a_S f(a_S)}\right)^2+4 \, \overline{a_S^2 f(a_S)}}},
\label{eq:sick_gen}
\end{equation}
which corresponds to Eq. \eqref{eq:rc}, for $r=\lambda \overline{a_S}/\mu$. The fraction of sick-leaving nodes is $n_p= \int da_S \rho_S(a_S) (1-f(a_S))$.

Now we consider the \textit{uniform strategy}, with $f(a_S)=(1-p)$ $\forall a_S$: in this case the fraction of sick-leaving nodes is $n_{p}^U=p$. Moreover:
\begin{equation}
\overline{a_S f(a_S)}= \int da_S \rho_S(a_S) a_S (1-p) = (1-p) \overline{a_S},
\label{eq:aI_U}
\end{equation}
\begin{equation}
\overline{a_S^2 f(a_S)}= \int da_S \rho_S(a_S) a_S^2 (1-p) = (1-p) \overline{a_S^2}.
\label{eq:aSaI_U}
\end{equation}
Therefore, by replacing Eqs. \eqref{eq:aI_U}-\eqref{eq:aSaI_U} into Eq. \eqref{eq:sick_gen}, we obtain Eq. \eqref{eq:rcU} for the epidemic threshold of the uniform strategy.

Let us now consider the \textit{targeted strategy}, where $f(a_S)=\theta(a^*-a_S)$ (with $\theta$ the Heaviside function and $a^* \in [a_m, \eta a_m]$). In order to compare the different strategies, we fix $a^*$, i.e. $y=a^*/a_m$, so that the fraction of sick-leaving nodes is equal to $p$, as in the uniform strategy. The fraction of sick-leaving nodes is:
\begin{equation}
n_{p}^T= \int da_S \rho_S(a_S) (1- \theta(a^*-a_S))= \frac{y^{-\nu}-\eta^{-\nu}}{1-\eta^{-\nu}}.
\end{equation}
Therefore, we impose $n_{p}^T=p$ and we obtain:
\begin{equation}
y=(p + \eta^{-\nu} (1-p))^{-1/\nu}.
\end{equation}
This relation must respect that $a^* \in [a_m,\eta a_m]$, i.e. $y \in [1,\eta]$: this is always guaranteed since $p \in [0,1]$, $\eta \in (1,\infty)$ and $\nu \geq 0$. Thus, for this strategy:
\begin{equation}
\overline{a_S f(a_S)}= \int da_S \rho_S(a_S) a_S \theta(a^*-a_S) = \overline{a_S} \frac{1-y^{1-\nu}}{1-\eta^{1-\nu}},
\label{eq:aI_T}
\end{equation}
\begin{equation}
\overline{a_S^2 f(a_S)}= \int da_S \rho_S(a_S) a_S^2 \theta(a^*-a_S) = \overline{a_S^2} \frac{1-y^{2-\nu}}{1-\eta^{2-\nu}}.
\label{eq:aSaI_T}
\end{equation}
Then, by replacing Eqs. \eqref{eq:aI_T}-\eqref{eq:aSaI_T} into Eq. \eqref{eq:sick_gen}, we obtain Eq. \eqref{eq:rcT} for the epidemic threshold of the targeted strategy.

Finally, we focus on the $\varepsilon$\textit{-targeted strategy}, $f(a_S)=\theta(a^*-a_S)+\varepsilon \theta(a_S-a^*)$ (where $\theta$ is the Heaviside function, $a^* \in [a_m,\eta a_m]$ and $\varepsilon \in [0,1]$). Also here, in order to compare the different strategies, we fix $a^*$, i.e. $y=a^*/a_m$, so that the fraction of sick-leaving nodes is $p$ as in the other strategies. The fraction of sick-leaving nodes is:
\begin{equation}
\begin{aligned}
n_{p}^{\varepsilon}&= \int da_S \rho_S(a_S) (1- \theta(a^*-a_S)- \varepsilon \theta(a_S-a^*))\\
&= (1-\varepsilon) \frac{y^{-\nu}-\eta^{-\nu}}{1-\eta^{-\nu}}.
\end{aligned}
\end{equation}
Therefore, by imposing $n_{p}^{\varepsilon}=p$ we obtain:
\begin{equation}
y=\left(\eta^{-\nu} + p \frac{1 - \eta^{-\nu}}{1-\varepsilon} \right)^{-1/\nu}.
\end{equation}
By imposing that $a^* \in [a_m, \eta a_m]$, i.e. $y \in [1,\eta]$, we obtain a condition on $\varepsilon$: $\varepsilon \leq 1- p$, otherwise $a^*$ would have unacceptable values. Thus, for this strategy:
\begin{equation}
\begin{aligned}
\overline{a_S f(a_S)}&= \int da_S \rho_S(a_S) a_S (\theta(a^*-a_S)+ \varepsilon \theta(a_S-a^*)) \\
&= \overline{a_S} \frac{1-y^{1-\nu}+\varepsilon (y^{1-\nu}-\eta^{1-\nu})}{1-\eta^{1-\nu}},
\end{aligned}
\label{eq:aI_E}
\end{equation}
\begin{equation}
\begin{aligned}
\overline{a_S^2 f(a_S)}&= \int da_S \rho_S(a_S) a_S^2 (\theta(a^*-a_S)+ \varepsilon \theta(a_S-a^*))\\
&= \overline{a_S^2} \frac{1-y^{2-\nu}+\varepsilon (y^{2-\nu}-\eta^{2-\nu})}{1-\eta^{2-\nu}}.
\end{aligned}
\label{eq:aSaI_E}
\end{equation}
Then, by replacing Eqs. \eqref{eq:aI_E}-\eqref{eq:aSaI_E} into Eq. \eqref{eq:sick_gen}, we obtain Eq. \eqref{eq:rcE} for the epidemic threshold of the $\varepsilon$-targeted strategy.

\section*{Appendix C: Sketch of the Gillespie-like algorithm for numerical simulations}
\label{app:C}
In this Appendix we present schematically the Gillespie-like algorithm \cite{gillespie1976general,mancastroppa2019burstiness,mancastroppa2020active} implemented to numerically simulate the SIR dynamics. 

We consider a population of $N$ nodes and we assign to each node their activities $(a_S,a_I,a_R)$ extracted from the distribution $\rho(a_S,a_I,a_R)$, which is fixed by the specific adaptive behavior considered. We let the network evolve without the epidemic process, i.e. in the absorbing state, up to a relaxation time $t_0$ ensuring that the activation dynamics of the network relaxes to equilibrium:
\begin{enumerate}
\item at time $t=0$, the first activation time $t_i$ for all the nodes $i$ is extracted from their distribution $\Psi_i(t_i)=a_S^i e^{-a_S^i t_i}$;
\item \label{ii} the node $i$ with the lowest activation time $t_i$ activates, creates a link with a node selected uniformly at random in the population and the new activation time of $i$ is set to $t_i+ \tau$, with $\tau$ drawn from the inter-event time distribution $\Psi_i(\tau)=a_S^i e^{-a_S^i \tau}$;
\item the generated link is removed and the process is iterated from the point \ref{ii}, up to time $t_0$.
\end{enumerate}
At time $t=t_0$, the network activation dynamics has relaxed to equilibrium, each node has an activation time $t_i>t$ and the epidemic dynamics is introduced:
\begin{enumerate}
\item the entire population is initialized in the susceptible status ($S$) except for the node with the highest $a_I^i$ which is set as the infection seed ($I$) \cite{nature2013boguna};
\item \label{II} the node $i$ with the lowest $t_i$ activates and the current time is set to $t=t_i$. The nodes infected at time $t'$ recover at time $t_i$ with probability $1-e^{-\mu (t_i-t')}$: recovered nodes change their activity $a_I \to a_R$; 
\item the active node $i$ creates a link with a node selected uniformly at random in the population. If the link connects a susceptible $S$ and an infected node $I$, the susceptible node gets infected with probability $\lambda$ (infectious contact) and the new infected node changes its activity $a_S \to a_I$; otherwise, nothing happens during the contact;
\item the new activation time of the active node $i$ is set to $t_i + \tau$, with $\tau$ drawn from the inter-event time distribution $\Psi_i(\tau)=a_k^i e^{-a_k^i \tau}$, where $a_k^i$ denotes the node current activity, depending on its health status $k$;
\item the generated link is removed and the process is iterated from point \ref{II}, until the system reaches the absorbing state with no infected nodes.
\end{enumerate}

\bibliographystyle{naturemag}
\bibliography{BiblioAdaptive}

\end{document}